\documentclass[traditabstract]{aa}
\usepackage{graphicx}
\usepackage{txfonts}
\usepackage{amssymb}
\usepackage{lscape}

\begin{document}

\def\mpc{h_{75}^{-1} {\rm{Mpc}}} 
\def\kpc{h_{75}^{-1} {\rm{kpc}}}
\newcommand{\mincir}{\raise
-2.truept\hbox{\rlap{\hbox{$\sim$}}\raise5.truept\hbox{$<$}\ }}
\newcommand{\magcir}{\raise
-2.truept\hbox{\rlap{\hbox{$\sim$}}\raise5.truept\hbox{$>$}\ }}
\newcommand{\ha}{\mathrm{H}\alpha}
\newcommand{\hb}{\mathrm{H}\beta}

\title{X-ray AGN in the XMM-LSS galaxy clusters:
\\ 
no evidence of AGN suppression.}

\author{E. Koulouridis\inst{1}, M. Plionis\inst{2,3,1}, O. Melnyk\inst{4,5}, A.
Elyiv\inst{4,6,7},
  I. Georgantopoulos\inst{1}, N. Clerc\inst{8}, J. Surdej\inst{4}, L.
Chiappetti\inst{9}, M. Pierre\inst{10}}

\institute{Institute for Astronomy \& Astrophysics, Space Applications \&
  Remote Sensing, 
National Observatory of Athens, Palaia Penteli 15236, Athens, Greece.
\and Physics Department of Aristotle University of Thessaloniki,
University Campus, 54124, Thessaloniki, Greece
\and Instituto Nacional de Astrof\'{\i}sica Optica y Electr\'onica, Puebla,
C.P. 72840, M\'exico.
\and Institut d'Astrophysique et de G\'eophysique, Universit\'e de Li\`ege, 4000
Li\`ege, Belgium
\and Astronomical Observatory, Taras Shevchenko National University of Kyiv, 3
Observatorna St., 04053 Kyiv, Ukraine
\and Dipartimento di Fisica e Astronomia, Universit\`a di Bologna, Viale Berti
Pichat 6/2, I-40127  Bologna, Italy
\and Main Astronomical Observatory, Academy of Sciences of Ukraine, 27 Akademika
Zabolotnoho St., 03680 Kyiv, Ukraine
\and Max-Planck-Institute for Extraterrestrial Physics, Giessenbachstrasse 1,
85748, Garching, Germany
\and INAF, IASF Milano, via Bassini 15, I-20133 Milano, Italy
\and Service d'Astrophysique, AIM, CEA Saclay, F-9191 Gif sur Yvette.}

\date{\today}

\abstract{We present a study of the overdensity of X-ray-selected active
galactic nuclei (AGN) in 33 galaxy clusters
in the XMM-LSS field (The XMM-Newton Large Scale Structure Survey), up to
redshift $z=1.05$ and further divided into a lower 
($0.14\leq z\leq 0.35$) and a higher redshift ($0.43\leq z\leq1.05$) subsample.
Previous studies have shown that the presence of X-ray-selected AGN in rich
galaxy clusters is suppressed, since 
their number is significantly lower than what is expected from the high galaxy
overdensities in the area. In the 
current study we have investigated the occurrence of X-ray-selected AGN in low
($\langle L_x, bol\rangle=2.7\times10^{43}$ erg/s) and moderate 
($\langle L_x, bol\rangle=2.4\times10^{44}$ erg/s) X-ray luminosity galaxy
clusters 
in an attempt to trace back the relation between high-density environments and
nuclear activity.
Owing to the wide contiguous XMM-LSS survey area, we were able to extend the
study to the cluster outskirts. We 
therefore determined the projected overdensity of X-ray point-like sources
around each cluster out to $6r_{500}$ radius, 
within $\delta r_{500}=1$ annulus, with respect to the field expectations based
on the X-ray source $\log N -\log S$ of 
the XMM-LSS field.
To provide robust statistical results we also conducted a consistent stacking
analysis 
separately for the two $z$ ranges. We investigated whether the observed X-ray
overdensities are to be expected thanks to the 
obvious enhancement of galaxy numbers in the cluster environment by also
estimating the corresponding optical galaxy 
overdensities, and we assessed the possible enhancement or suppression of AGN
activity in clusters.

We find a positive X-ray projected overdensity in both redshift ranges at the
first radial bin, which however has the same 
amplitude as that of optical galaxies. Therefore, no suppression (or
enhancement) of X-ray AGN activity with respect 
to the field is found, in sharp contrast to previous results based on rich
galaxy clusters, implying that the mechanisms 
responsible for the suppression are not as effective in lower density
environments. After a drop to roughly the background level 
between 2 and $3 r_{500}$, the X-ray overdensity exhibits a rise at larger
radii, significantly greater than the corresponding optical 
overdensity. The radial distance of this overdensity \textquotedblleft bump",
corresponding to $\sim 1.5 - 3$ Mpc, depends on the richness of 
the clusters, as well as on the overall X-ray overdensity profile.

Finally, using the redshift information, photometric or spectroscopic, of the
optical counterparts, we derive the 
spatial overdensity profile of the clusters. We find that the agreement between
X-ray and optical overdensities in the 
first radial bin is also suggested in the 3-dimensional analysis. However,
we argue that the X-ray overdensity \textquotedblleft bump" at larger radial
distance is at least partially a result of flux 
boosting by gravitational lensing of background QSOs, confirming previous
results. For high-redshift clusters the enhancement of 
X-ray AGN activity in their outskirts appears to be intrinsic.
We argue that a spatial analysis is crucial for disentangling irrelevant
phenomena affecting the projected analysis, 
but we are still not able to report statistically significant results on the
spatial overdensity of AGN in clusters or their 
outskirts because we lack the necessary numbers.}

\authorrunning{E. Koulouridis et al.}
\titlerunning{X-ray AGN in the XMM-LSS clusters}
\maketitle

\section{Introduction}

As one of the most powerful extragalactic phenomena,
active galactic nuclei (AGN) are a valuable tool in the study of the universe,
since they can be 
used as cosmological probes, provide answers to various problems 
of galaxy evolution, and shed light on the innermost regions of galaxies, 
where the super massive black hole resides and highly energetic processes take
place.
However, the triggering mechanism of the omnipresent, but not always active,
black hole 
is still elusive. Although major merging of gas-rich galaxies seems to be a 
highly probable mechanism for the triggering of nuclear activity (e.g., Sanders
et al. 1988; Barnes \& Hernquist
1991; Hopkins et al. 2006), recent studies of the morphology of AGN hosts find
no evidence of a major merging-AGN 
connection (e.g., Cisternas et al. 2011; Kocevski et al. 2012).
On the other hand, minor merging and interactions are still strongly disputed,
while secular evolution also seems able 
to feed the central engine since many AGN are found to be isolated and
undisturbed (e.g., Hopkins \& Hernquist 2006; 
Cisternas et al. 2011; Kocevski et al. 2012).
Therefore, the effect of the environment on the activity of the nucleus
and vice versa is still fairly undetermined, but nevertheless crucial. 
Galaxy clusters represent the one end of the 
density spectrum in our universe, and as such it is an ideal place to
investigate 
the effects of dense environment in the triggering of AGN, especially since 
an excessive number of X-ray point-like sources are undoubtedly found there
(e.g., Cappi et al. 2001; Molnar et al. 2002; Johnson et al. 2003; D'Elia et al.
2004; Gilmour et
al. 2009). Specifically, for the XMM-LSS field investigated in the current
study, Melnyk et al. (2013) have found that 60\% 
of X-ray-selected AGN reside in the overdense regions of group-like environment.

Theoretically the feeding of the black hole can only be achieved by means of a 
non-axisymmetric perturbation that induces mass inflow. This kind of 
perturbation can be provided by interactions and merging between two galaxies,
and the result of the
inflow is the feeding of the black hole and activation of the AGN
phase (e.g., Umemura 1998; Kawakatu et al. 2006; Koulouridis et al. 2006a,
2006b, 2013, Ellison et al. 2011;
Silverman et al. 2011; Villforth et al. 2012; 
Hopkins \& Quataert 2011). Thus, the cluster environment, where 
the concentration of galaxies is very high relative to the field, would also
seem favorable to
AGN. However, the rather extreme 
conditions within the gravitational potential of a galaxy cluster can work in
the opposite 
direction as well. Ram pressure from the inter cluster medium
(henceforth ICM) able to strip/evaporate the cold gas reservoir of galaxies
(Gunn \& Gott 1972;
Cowie \& Songaila 1977; Giovanelli et al. 1985) can strongly affect
the feeding of the AGN. Nevertheless, other studies 
have argued that ram pressure stripping cannot be as effective
in transforming blue sequence galaxies to red
(e.g., Larson et al. 1980; Balogh et al. 2000, 2002;
Bekki et al. 2002; van den Bosch et al. 2008; Wetzel et al. 2012), especially
in 
lower density clusters where other processes should take place as well. 
Large velocity dispersion of
galaxies within clusters could also prevent the effective interactions (Aarseth
\& Fall 1980),
particularly mergers, while the fast \textquotedblleft grazing"
bypassing galaxies 
may also cause gas stripping by \textquotedblleft harassment"
(e.g., Natarajan et al. 2002, Cypriano et al. 2006).
However, the efficiency of this phenomenon has once more been
questioned (e.g., Giovanardi et al. 1983). A combination 
of the above mechanisms, in addition to the possible prevention of 
accretion of halo mass into cluster galaxies (\textquotedblleft strangulation";
e.g., Larson
et al. 1980; Bekki et al. 2002; Tanaka et al. 2004) may, in fact, suppress the
AGN activity in clusters
despite the number of potentially merging and interactive galaxies.

When using only optical data, the results seem to remain inconclusive.
Early studies reported that AGN are less 
frequent in galaxy clusters than in the field (Osterbrock 1960; Gisler
1978; Dressler, Thompson \& Schectman 1985) and recent large-area surveys
support this suggestion (Kauffmann et al. 2004; Popesso \& Biviano
2006; von der Linden et al. 2010; Pimbblet et al. 2013). Other studies,
however, have found no differences between clusters and field (e.g., Miller
et al. 2003), at least when selecting the weak AGN (e.g., Martini et
al. 2002; Best et al. 2005; Martini et al. 2006; Haggard et
al. 2010). We should note here that considering only the optical
wavelengths is not the optimal way of finding AGN since they suffer
greatly from absorption. Especially if gas depletion is at play and
low accretion rates are expected, then  most of the spectral
signatures of the AGN could be \textquotedblleft buried" in the host galaxy.

Radio-loud AGN on the other hand, seem to be more clustered than any
other type of galaxy (Hart, Stocke \& Hallman 2009) and are often
associated with BCGs (brightest cluster galaxies)(e.g., Best 2004; Best
et al. 2007). In addition, 
the fraction of X-ray AGN in BCGs is higher than in other cluster galaxies 
(e.g., Hlavacek-Larrondo et al. 2013, and references therein).
These findings can be attributed to hot gas accretion from the hot X-ray cluster
halo, 
although gas from any other source fueling the black hole at low accretion rates
would also have the same effect. If the hot gas accretion is a possible fueling
mechanism for the X-ray 
AGN, as well, then we should expect them to reside primarily within clusters.

Undoubtedly, the best way to detect active galaxies is through X-ray
observations (e.g., Brandt \& Alexander 2010). 
However, during the previous decade, only a small fraction of X-ray point-like
sources in clusters
had positive confirmation as true cluster members (see Martini et
al. 2002; Davis et al. 2003; Finoguenov et al. 2004; Arnold et al. 2009),
leaving the
question of whether the positive X-ray overdensities found in galaxy clusters
represent enhancement 
or suppression of the nuclear activity unanswered. More recent 
studies, however, report more conclusive results by comparing X-ray to optical
data. Koulouridis \& Plionis (2010) demonstrate the
significant suppression of X-ray-selected AGN in 16 rich Abell clusters (Abell
et al. 1958) by comparing
the X-ray point source overdensity to the optical galaxy
overdensity. Ehlert et al. (2013; 2014) argue that the X-ray AGN fraction
in the central regions of 42 of the most massive clusters known
to date is about three times lower than the field value using the same
technique. 
More importantly, after having complete
spectroscopy for their X-ray point source sample, Haines et al. (2012) argue
that X-ray AGN
found in massive clusters are an infalling population, which is
\textquotedblleft extinguished" later, and confirm the suppression in rich
clusters. On the other hand, Martini et al. (2013) argue that this
trend is not confirmed for a sample of high-redshift clusters
($1.0<z<1.5$), where the presence of luminous X-ray AGN ($L_{(0.5-2{\rm\;
keV})}>10^{43}$ erg/s) is consistent with the field.
We note, however, that the high-redshift regime studied and the large AGN
photometric
redshift uncertainties ($\sigma_z=0.12(1+z)$, double that of normal galaxies)
introduce some level
of uncertainty to the results.
Neverteless, they agree with findings from the DEEP2 Redshift
Survey\footnote{http://deep.ps.uci.edu/} 
that show that only below $z$=1.3 does the fraction of blue galaxies
in groups drop rapidly and become constant below $z$=1 (Gerke et al. 2007),
while 
the red fraction correlates weakly with overdensity above $z$=1.3 (Cooper et al.
2007). 
In the present study we only deal with clusters $z<$1.05, where the cluster's
population is dominated by early type red galaxies. 
Finally, we
should also mention that an indirect way to address the issue is by
X-ray clustering analyses, but still their results also remain 
inconclusive (see relevant discussion in Haines et al. 2012 \S5.2).

Considering the above, there is still the need to clarify the
influence of the environment on the AGN phenomenon. And while the majority of
the 
above studies are dealing with the most massive and rich clusters, the 
population of moderate-to-poor clusters is still overlooked.\footnote{We should 
note that the categorization of galaxy clusters to different richness classes is
not explicit, and 
it is safer to be used statistically. 
Nevertheless, the relation between mass and X-ray 
properties is well studied (e.g., Edge \& Stewart 1991a, b;
Finoguenov, Reiprich \& Bohringer, 2001) and, also considering more recent
studies of X-ray luminous clusters (e.g., Ebeling et al. 2010), 
we can infer that massive clusters have X-ray luminosities higher than
$\sim5\times10^{44}$ erg/s and  
temperatures higher than $kT>5$ keV.} If the 
reason for the deficiency of X-ray AGN in rich clusters is the strong
gravitational potential, which provides the necessary conditions for
the suppression (whichever these may be: gas stripping, 
strangulation, tidal stripping, evaporation, high velocity dispersion,
etc.), one would expect 
the AGN presence to rise in shallower gravitational potentials.
A similar relation between the strength of the gravitational potential and 
star formation quenching (Popesso et al. 2012) supports the above 
expectation (see also Wetzel et al. 2012). 
Another issue is the radial extent of the search 
for X-ray AGN around clusters. An enhancement of AGN activity is observed
far from the cluster's center (e.g., Fassbender et al. 2012),
and it could be due to an infalling population (Haines et al. 2012)
coming from the \textquotedblleft outskirts" of the clusters where the
concentration of galaxies is still high. The question is where should
we place the \textquotedblleft outskirts" and to what extent. 
Most studies could not reach farther than a 2$r_{500}$ radius, although
the overdensity profile of optical galaxies remains 
higher than the field level even beyond that radius (e.g., Ehlert et
al. 2013). Finally, what is also overlooked is
the background overdensity of X-ray sources in the area of
clusters. In Koulouridis et al. (2010), we used SDSS (Sloan Digital Sky Survey)
optical data for all 
the detected X-ray point-like sources within a 1 Mpc radius, and argue that
their positive overdensity values were associated with 
background QSOs rather than cluster members. A possible cause is the
gravitational lensing of background sources,
which is unimportant when compared to the large number of optical galaxies in
clusters but can become very important for X-ray sources and affect
the assessment of their clustering.

In the current study, our aim is to investigate the AGN phenomenon in
the environment of moderate and poor clusters
located in the XMM-LSS contiguous field of 11.1 deg$^2$. We
identify all possible X-ray AGN candidates, which we define as sources
with $L_{(0.5-2{\rm\; keV})}>10^{42}$ erg/s at the redshift of the cluster, and
compare their overdensity in the area of the clusters to 
the respective overdensity of optical galaxies, available by the CFHT
legacy survey. Such a large contiguous area gives us the unique
opportunity not only to use a large cluster sample but also to
extend our search for X-ray AGNs around clusters at great distances,
reaching homogeneously up to a 6$r_{500}$ radius. In
addition, we make use of photometric redshift data calculated
specifically for X-ray-selected AGN hosts, in an attempt to assess the true
number of
X-ray AGNs in our clusters and clarify the effect of the excessive
overdensity of background X-ray sources (e.g., Koulouridis \& Plionis
2010).

We describe our samples and methodology in \S2, while our results
and conclusions are presented in \S3 and \S4, respectively. Throughout
this paper we use $H_0=72$ km/s/Mpc, $\Omega_m=0.27$, and
$\Omega_{\Lambda}=0.73$.

\section{Sample selection \& methodology}

\subsection{The XMM-LSS}

The XMM-LSS field is part of the XXL survey, which is the largest international
XMM project approved to date ($>$6Msec), surveying 
two 25 deg$^2$ fields at a  depth of $\sim3\times 10^{-15}$ erg cm$^{-2}$
s$^{-1}$ 
in the [0.5-2] keV band. It
occupies an area of 11.1 deg$^2$ and is located
at high galactic latitudes (2$^{h}14^{m}<$ ra $<2^{h}30^{m}$;
-6$\raisebox{1ex}{\scriptsize o}25'<$ dec $<-2
\raisebox{1ex}{\scriptsize o}35'$, J2000.0, see also Fig. 1). 
It is contiguous, consists of 94 pointings with effective exposures
from 10 ks to 40 ks and contains the Subaru X-ray Deep Survey (SXDS; Ueda et al.
2008) that covers  
1.14 deg$^2$ of the area\footnote{The data are available in the Milan DB in
  the 2XLSSd and 2XLSSOPTd tables. See Chiappetti et al. (2013) for
  details.}. The XMM-LSS full-exposure
field contains 6342 sources$^2$, 5737 of them detected in
the soft [0.5-2] keV band down to a detection likelihood of 15 ($\sim1\%$
spurious, see Pacaud et al. (2006) for 
more details about the source detection and the statistics). The effective flux
limit in the soft band is 
F$_{(0.5-2 {\rm\; keV})}= 3\times10^{-15}$ erg s$^{-1}$ cm$^{-2}$, which
corresponds  
to 50\% sky coverage according to the effective area curves of Elyiv et al.
(2012). The large majority of the
X-ray sources is point-like ($N=5570$), which are mainly AGN since the expected
stellar contamination is near 3\%.
twenty-eight percent of the sources are excluded mostly due to the lack of
sufficient photometry bands ($<$4), lack
of counterpart ($\sim$4\%), and double-peaked photometric redshift solution. 

We therefore used 72\% of the full sample, of which 
17\% had spectroscopic redshifts (see \S2.6), although the 
respective completeness around our clusters is $\sim$81\%, and spectroscopic
redshifts 
are available for $\sim$24\% of the total (see \S3.3).  
respectively.
More details about the catalog's
sensitivity, the effective area curve, and the $\log N - \log S$ 
can be found in Elyiv et al. (2012) and in the multiwavelength catalog
description paper by Chiappetti et al. (2013). We note that the soft-band $\log
N - \log S$ 
used in the current study, calculated by Elyiv et al. (2012), is lower than 
those of the 2XMM (Ebrero et al. 2009) and
COSMOS (Cappelluti et al. 2007) surveys (with deviations not
exceeding the $2 - 3\sigma$ Poisson level), but in excellent
agreement with those of the XMM Medium Deep Survey (XMDS, Chiappetti et al.
2005).

\begin{figure}
\resizebox{9cm}{6cm}{\includegraphics{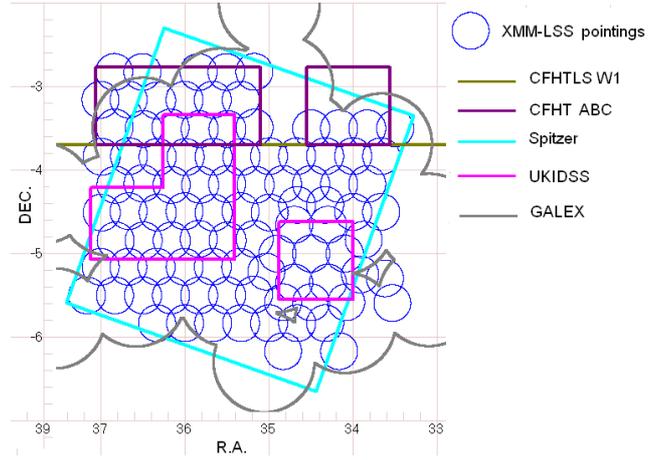}}
\caption{Multiwavelength coverage of the XMM-LSS field. Colored borders
demarcate the regions covered by
The Canada-France-Hawaii Telescope Legacy Survey (CFHTLS W1, below the green
horizontal line), the Canada-France-Hawai Telescope (CFHT) ABC
supplementary pointings, Spitzer Space Telescope (SST), the UKIRT Infrared Deep
Sky Survey (or UKIDSS) and 
The Galaxy Evolution Explorer (GALEX).}
\end{figure} 

\subsection{The cluster sample}

The clusters of galaxies used in this study 
were selected to meet the following conditions:
\begin{enumerate} 
\item They belong to the list of unambiguously confirmed C1 clusters
of the XMM-LSS field
(Adami et al. 2011; Pacaud et al. 2006; Pierre et al. 2006).
\item They are not located near the edges of the XMM-LSS field, to ensure
the complete detection of all point-like sources within the radii of interest.
\item They belong to the redshift range $z<1.05$. 
The upper limit is set at a reasonable redshift
above which the lower X-ray source luminosity that corresponds 
to the lower flux limit is becoming very high
(the lower flux limit is set at a
certain value; see next paragraph). 
\end{enumerate}
For all clusters that meet the above criteria, we list in
Tables 1 and 2 (from Clerc et al. 2014, in prep.):
the temperatures and X-ray luminosities (giving an estimate of their
richness and of the depth of the gravitational potential of the
cluster), and the $r_{500}$ radius calculated by fitting a beta model 
to the extended emission (see also Pacaud et al. 2007).
The positions of all the clusters and their corresponding 4$r_{500}$ 
(5$r_{500}$ for the high-$z$ sample)
radius are overplotted on the XMM-LSS X-ray map in Fig. 2. 
In the full XMM-LSS sample used in this work, the median number of photons used
for the X-ray spectral 
fit is 350 (in [0.5-2] keV), 
and the median statistical error on the temperatures is 15\%. Regarding the 
temperatures, the interested reader can refer to Willis et al. (2006) (in their
appendix),
where they simulated various clusters and applied a similar X-ray spectral
measurement procedure 
to derive the temperature uncertainties.

We limited our analysis to sources above a flux limit of
 $3\times 10^{-15}$ erg s$^{-1}$ cm$^{-2}$, since at
lower fluxes, sources were scarcely detected in the XMM-LSS survey
and thus
the resulting $\log N-\log S$ bears large uncertainties in this flux regime.

To have homogeneous samples in 
X-ray luminosity, as well as to study possible evolutionary effects,
we divided the 33 clusters into two subsamples, a
lower and higher redshift sample. 
The low-$z$ sample ($0.14\leq z\leq0.35$) consists of 19 clusters with average 
X-ray luminosity $\langle L_x, bol\rangle=2.7\times10^{43}$ erg/s and average
temperature $<kT>=2.0$ keV, while
the high-$z$ sample ($0.43\leq z\leq1.05$) of 14 clusters with average 
X-ray luminosity $\langle L_x, bol\rangle=2.4\times10^{44}$ erg/s and average
temperature $<kT>=3.1$ keV. Based on the 
X-ray luminosity and temperature of our clusters, we have no rich clusters 
in our samples (Alshino et al. 2010), with the exception of XLSSC 006, while the
low-$z$ sample mostly consists 
of poor systems and the high-$z$ mostly of intermediate systems. 
We imposed a limiting X-ray luminosity of $10^{42}$ erg/s for our
X-ray sources, in order to securely select X-ray AGN.
Therefore, the corresponding X-ray flux limit is such that for the low-$z$
sample, the rest-frame luminosity limit is always 
$L_{(0.5-2{\rm\; keV})}=10^{42}$ erg/s, which means that we are complete in
X-ray luminosity. However, this is not the
case for the high-$z$ sample. Especially for clusters above $z\sim0.95$ the
X-ray luminosity
limit is $L_{(0.5-2{\rm\; keV})}=1.7\times 10^{42}$ erg s$^{-1}$ and reaches to
$L_{(0.5-2{\rm\; keV})}=1.4\times 10^{43}$ erg s$^{-1}$ for the two clusters at
redshift
$z\sim1$ (see Table 2).

\begin{figure}
\resizebox{9cm}{9cm}{\includegraphics{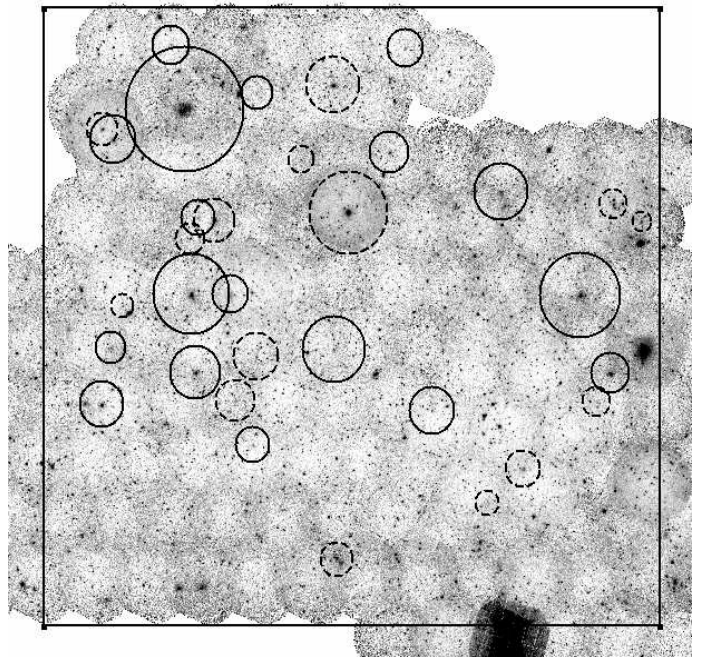}}
\caption{Position and 4$r_{500}$ (5$r_{500}$) radius of the current study's
  low-redshift (high-redshift - dashed circles) clusters in the XMM-LSS region
(square).}
\end{figure}

\subsection{X-ray source overdensity}
 The X-ray AGN overdensity, in a given area, is estimated according to
$$\Delta_x=\frac{N_x-N_{\rm exp}}{N_{\rm exp}}$$ 
where $N_x$ the
number of X-ray point-like sources detected in the area and $N_{\rm exp}$
is the expected number according to the $\log N-\log S$ within the same
area. 
To calculate the value of $N_x$, we identify all point-like sources located
within five (six for the high-$z$ clusters) 
radial annuli between $n$ and
$(n+1) r_{500}$, where $n=$0,1,2,..5. We consider only the sources 
with X-ray fluxes $f_{(0.5-2{\rm\; keV})}>f_{\rm lim}$, where $f_{\rm lim}$ is
the specific value
of the flux for which the AGN X-ray luminosity at the distance of
any cluster is $L_{(0.5-2{\rm\; keV})}=10^{42}$erg/s.
The large contiguous area of the XMM-LSS survey allows us to expand
our search for X-ray AGN activity at large radii from the cluster 
cores and gives us the opportunity to explore possible evolutionary
trends of nuclear activity as galaxies enter the cluster's gravitational
potential from their outskirts.

Because of the cluster's diffuse X-ray emission, in most cases we may
not be able to sufficiently resolve the central region and could
possibly fail to detect X-ray AGN in that area. Thus, we choose to exclude
a fraction of the first $r_{500}$ annulus from each cluster to avoid
introducing a possible bias into our results. 
However, the area that has to be excluded depends on the strength of
the extended emission, as well as on the imposed lower flux limit of
the X-ray sources we wish to detect, and therefore
it varies from cluster to cluster.
In Koulouridis et al. (2010), we excluded the regions with $r<r_{\rm
  core}$, where $r_{\rm core}$ was found by fitting a $\beta$ model 
to the extended X-ray emission.
For the current analysis, 
we inspected all clusters
visually and chose to exclude the inner 0.5$\times
r_{500}$ homogeneously from all clusters. We found that the extracted
area eliminates the problem of diffuse X-ray emission in all clusters
and furthermore allows us to do a meaningful comparison and stacking of the
first annuli
of different clusters.  

To calculate the expected number $N_{\rm exp}$ of X-ray sources in the
field, we followed the procedure described below,
considering each time the same area of the detector and 
the same characteristics of the actual observation: 
\begin{enumerate} 
\item From the $\log N-\log S$ of the XMM-LSS survey (see Elyiv et al. 2012 for
a more detailed analysis), 
we derive the total number ($N_f$) of expected sources in the area per flux bin.
\item We consider $1000\times N_f$ sources with random fluxes within
  the flux range of each bin and random position within the area of
  interest.
\item We derive the probability $P_i$ that the source $N_{fi}$ is
  actually detected in the specific area of the detector. The probability
  is a function of the off-axis position (vignetting), background, and 
  exposure time.
\item We calculate the sum: $\sum_{j=n}^{53}\sum_{i=1}^{1000N_f}
  N_{fi}\times P_i/1000$, which gives us the total number, $N_{\rm exp}$, 
of expected X-ray sources that have fluxes above the respective value
of the $n^{th}$ bin of the $\log N-\log S$, where the total number of bins is
53.
\end{enumerate}

To validate the above procedure, we randomly picked ten clusters
with $~200000$ random sources,
and we attempted to reproduce the $\log N-\log S$ by using the area curve of
the XMM-LSS field. To this end, we binned the sources in the same flux
bins and then divided the number of sources in each bin 
with the fraction of the XMM-LSS area where we could actually detect
them. We indeed recover the input $\log N-\log S$ with great
accuracy as we can see in Fig. 3. The apparent deficiency of sources in 
the flux bins above $10^{-13}$erg s$^{-1}$ cm$^{-2}$, which is due to the 
rarity of these sources in the full XMM-LSS field, is very small ($<0.2$
sources/deg$^2$),
well within the errors, and not important to the current analysis.

\begin{figure}
\resizebox{9cm}{9cm}{\includegraphics{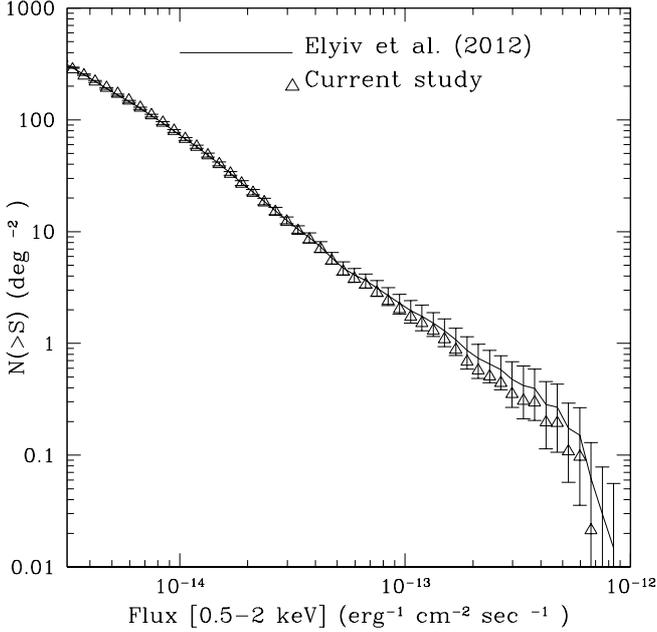}}
\caption{$\log N -\log S$ distribution in the soft band for the whole XMM-LSS 
sample (line). The triangles denote the test values derived from random sources
in the area 
of 10 galaxy clusters in the current study. Errorbars denote the 1$\sigma$
Poissonian uncertainty
of the Elyiv et al. (2012) points.}
\end{figure}

\subsection{Optical galaxy overdensity}

For calculating the optical overdensity in each
XMM-LSS cluster, we use the 
$i$-band magnitude of the CFHTLS, which is complete up to $m_i\sim 24$, a
crucial feature for our high
redshift clusters.

As for the X-ray overdensity, the relevant expression is 
$$\Delta_o=\frac{N_o-N_{o, \rm exp}}{N_{o, \rm exp}}\;,$$
where $N_o$ is the number of optical sources found in the area and
$N_{o, \rm exp}$ the expected background number within the same area. 
For the calculation of the galaxy density, within the regions
previously defined for the X-ray analysis, we include only those
sources with $i$-band magnitude that satisfy the criterion
$m_i^*-1<m_i<m_i^*+1$, where $m_i^*$ is the apparent $i$-band magnitude
that corresponds to the break of the luminosity function at the
redshift of each cluster. We should note that we have also conducted
the same analysis using  $m_i^*-0.5<m_i<m_i^*+0.5$, but the results
are exactly the same. For calculating $m*$, we used the 
K-correction values derived specifically for the CFHT $i$-band
magnitudes by Ramos et al. (2011) for elliptical galaxies,
since most of the cluster's galaxies are expected to be ellipticals.  
The background galaxy density is calculated from a 2 deg$^2$ field within the
XMM-LSS area.

We should note here that the comparison of the optical galaxy overdensity to
the X-ray AGN overdensity should not be considered explicit but rather
instructive. Although by using the $L_{(0.5-2{\rm\; keV})}>10^{42}$ erg/s
criterion, all X-ray
sources at the cluster rest frame are potential AGN, and
on the other hand, by using the $m_i^*-1<m_i<m_i^*+1$
criterion we choose the bright cluster
galaxies that could potentially host the AGN, 
there is no way of knowing if we are exactly probing 
 the same population of galaxies.

However, the method that we are using in the present
study and which has been used in various others (e.g., Martini et
al. 2013; Ehlert et al. 2014) over the past years with small
variations, can give us valuable information about the presence of AGN
in clusters. 
Finally, as an extra step, we also use the available
photometric and spectroscopic redshift data, in an attempt to
constrain the true overdensities of X-ray AGN better, since there
is evidence that their overdensities, within a $\sim$1
Mpc radius around rich clusters, may be affected by gravitational lensing
(Koulouridis et al. 2010). 

The use of the CFHTLS survey has two caveats 
related to our current study: it has \textquotedblleft holes" with no available
data 
as the result of star masking, and more importantly it does not
cover the whole XMM-LSS field. The former may cause the
underestimation of some cluster's optical overdensities, so 
we proceeded with corrections when necessary and we verified that, 
overall, the \textquotedblleft holes" do not affect our
statistical results based on stacking.
On the other hand, the region outside the CFHT Legacy Survey is instead covered 
by the CFHT ABC supplementary pointings. However, the available optical CFHT
photometric
bands in these fields are less than in the CFHTLS (one to three instead of five
bands), resulting
in less reliable photometric redshifts of the X-ray point source counterparts
(calculated using only four to six bands
depending on the Spitzer and UKIDSS coverage, see Fig.1). 
More importantly, there is no i band, which is essential for our projected
analysis, 
and no photometric redshifts of the optical galaxies, which are necessary for
our spatial analysis. 
Thus, when the ABC fields are included in our analysis, the results are given
separately and 
are treated with caution.

\subsection{Weighting and stacking}

To have more robust results, we also stack, at rest frame, the
respective annuli of all clusters (high-$z$ and low-$z$ always
separately). 
However, stacking the X-ray sources found in different pointings
of the XMM-LSS field is only possible if we eliminate the systematic
differences in the exposure time, background, and off-axis distance. To this
end, we multiply the sources found within 
a certain annulus in a cluster by a weight $w_{ij}$, where $i$ is the
number of the annulus and $j$ refers to the cluster, to eliminate the above
differences. 
In the case of the low-$z$ sample, the weight of an annulus $i$ of a cluster $j$
is
\[w_{ij}=N_{f\;ij}/N_{\rm exp\;\it ij}\;,\] where $N_{fij}$ is the number of
sources in the annulus $i$ 
of cluster $j$, calculated
directly from the $logN - logS$ before applying the corrective
factors of steps 2-4 of \S2.2, and thus it does not include any information
about 
the exposure time, background, and off-axis distance, whereas $N_{\rm exp\;\it
ij}$ 
is the number of expected sources that does 
include these corrections (see \S2.2).
We stress that 
the weighting is essential because we want to estimate the overdensity of the
merged annuli by the formula:
\[\Delta_i=\left(\sum^{K}_{j=1}N_{ij}\middle/\sum^{K}_{j=1}N_{\rm exp\,\it
ij}\right)-1\;.\]
where $N_{ij}$ is the number of detected x-ray sources and $N_{\rm exp\,\it ij}$
as before. 
The outcome of the formula is not 
the average overdensity of the clusters, since we are not adding the individual
cluster overdensities, 
but rather the total overdensity of the stacked regions, as we are summing the
number of sources of all clusters.
Consequently, we calculate the total X-ray source
merged overdensity in the annulus $i$ by the formula:
\[\Delta_{i}=\left(\sum^{K}_{j=1}N_{ij}w_{ij}\middle/\sum^{K}_{j=1}N_{\rm
exp\;\it ij}w_{ij}\right)-1\;,\]
where $K$ is the number of clusters. 
This procedure corrects the number of observed sources to the
number that we would have found in any certain annulus within the
XMM-LSS field if not for the differences in exposure time, background
and off-axis distance.

For the stacking analysis of the high-$z$ sample clusters we need to add
an extra correction, which occurs because the lower flux limit is now fixed to
the survey's lower limit  
of $3\times 10^{-15}$ erg s$^{-1}$ cm$^{-2}$ (see \S2.1) independently of the
redshift. 
This flux corresponds to a 
different limiting $L_{x, \rm lim}$ in each cluster.
Thus, as we go to higher redshift, the flux limit remains the same 
but the effective area under
investigation for a given $r_{500}$ radius is getting
smaller resulting in a progressively smaller number 
of expected or actually detected sources.
For example, considering the same cluster at two different redshifts, the
projected $r_{500}$ radius in arcmin is larger at the lower redshift. With a
fixed 
lower flux limit, the number of expected and detected sources in the larger area
at the
lower redshift is larger, rendering the low-redshift cluster 
\textquotedblleft heavier" in the stacking procedure, since it contributes more
in the sum of the sources.
To minimize this effect, we choose a random cluster, we set it as reference, 
and we normalize the area of every sample cluster to the respective one of the
reference cluster. The normalization should not eliminate the
intrinsic $r_{500}$ area differences, i.e. should correct only for 
the redshift difference of the clusters. Therefore, the weight formula is now
transformed to
\[w_{ij}=\frac{N_{f\;ij}}{N_{\rm exp\;\it
ij}}\times\frac{A_{ref}}{A_j}\times\frac{B_j}{B_{ref}}\;,\]
where $A_{ref}$ and $A_j$ are the projected reference and cluster area in
deg$^2$, respectively, whereas 
$B_{ref}$ and $B_j$ are the intrinsic reference and cluster area in Mpc$^2$,
respectively. 
The second term normalizes all cluster areas to the reference area, 
while the third enforces the intrinsic differences of $r_{500}$ between
different clusters. 
In that way we add all sources found above the constant
lower flux limit, within the respective $r_{500}$ annuli of any cluster, free
of the progressive area-diminishing effects.   
Consequently, as already mentioned, 
the stacking for the high-$z$ clusters is conducted between a range of
different limiting X-ray AGN luminosities ($\sim 1.7\times 10^{42} - 1.4\times
10^{43}$
erg/s), although this range is not that wide. This caveat is also present in
most other
studies that use a fixed lower flux limit and not a luminosity one (Haines et
al. 2012; 
Ehlert et al. 2014; Fassbender et al. 2012; Martini et al. 2013). The comparison
between
these studies is even more difficult due to the different redshift regimes 
under investigation, leading to a progressive loss of the less powerful AGN
population with
distance. However, although a direct comparison
between the low and high-redshift clusters is not possible in the
current study or between other studies, 
sample in two),
pointing out differences and similarities can still be useful.

Finally, we calculate the Poissonian error on the weighted number $n$ of
events, given by the formula:
\[\sigma\left(\sum n\right)=\sigma\left(\sum w\right)=\left(\sum
{w_i^2}\right)^{1/2}\;. \]

The number of optical galaxies, in each annulus, does not have to
be corrected before stacking, since optical data in the CFHTLS do not suffer 
from any obvious incompleteness or any other selection effect up to
$i_{mag}\simeq24$.
This limit is adequate for the present study, since we only have to exclude one
cluster from the optical analysis, XLSSC 078.

\subsection{Spatial overdensity analysis using redshifts and visual inspection
of counterparts}

To interpret our results and attempt to understand the physical mechanisms
behind
the observed X-ray overdensities, we would ideally like to be able to place  
the candidate X-ray AGN in the cluster or its outskirts.
Spectroscopy, however, is only available for $\sim 24\%$ 
of our X-ray point-like sources (Table 1), and thus we investigate the rest of
the objects based 
on photometric redshifts and visual inspection. Calculation of the photometric
redshifts 
is described shortly in this section and in more detail in Melnyk et al.
(2013). 

For each X-ray source, we take only one best rank optical CFHTLS counterpart
into
account, based on its distance from the 
X-ray source and its relative brightness
(rank=0 for a single very reliable counterpart or 1 for 
a less reliable but preferred counterpart, see
Chiappetti et al. 2013 for details). All counterparts with less than four
available photometry 
bands were also discarded. We therefore only considered 4555 point-like X-ray
sources ($\sim72\%$ completeness), 4450 of which have spectroscopic or
photometric redshifts with $z>0$
(non-stars). The list of redshifts for all XMM-LSS sources can be found in Table
2 of Melnyk et
al. (2013). For the photo-z determination, the
LePhare\footnotemark\footnotetext{
http://www.cfht.hawaii.edu/~arnouts/LEPHARE/lephare.html}
public code (Arnouts et al. 1999, Ilbert et al. 2006) was used. The
accuracy\footnotemark\footnotetext{$\sigma_{\Delta
  z/(1+z_{sp})}$ according to Hoaglin et al. (1983)} of the photometric redshift
calculation is $\sigma_{\Delta z/(1+z_{sp})}$=0.076, with 22.6\% of them
outliers for the case of counterparts having at least four photometric bands.
The redshift probability distribution (PDZ) of $\sim41\%$ of the sources is
PDZ=100, meaning that the solution is unique
and highly probable. The bulk of our sources, with unavalaible spectroscopic
redshifts, are from the above subsample, or they have
photometric redshifts calculated with at least seven bands. The latter have
PDZ$<$100, meaning larger uncertainties, 
although the spectro-z to photo-z relation is similat to that of the PDZ=100
sources (see Melnyk et al. 2013). 
The LePhare code also indicates secondary solutions, but in this case the
sources were rejected. Double-peak solutions 
are produced when only a few photometry bands are available, and that happens
mostly in the ABC supplementary fields.

Any X-ray counterpart to be
considered as a cluster member should have its
spectroscopic redshift within $\pm 2000$km/s of the cluster redshift
$z_{cl}$, or photometric redshift $z_{ph}$ within $\sigma (1+z_{cl})$,
where $\sigma$=0.065 to 0.076 depending on the available photometry
bands that were used for the calculation and the redshift probability
distribution 
(Melnyk et al. 2013). 

For the CFHT optical galaxies we used the photometric redshifts of the
CFHTLS-T0007 W1 field (Ilbert et al. 2006 and Coupon et al. 2009)
computed for three to five optical bands. The accuracy is 0.031 at $i<21.5$ and
reaches
$\sigma_{\Delta z/(1+z_{sp})} \sim 0.066$ at $22.5<i<23.5$. The fraction of
outliers
increases from $\sim 2\%$ at $i<21.5$ to  $\sim$10 - 16\% at
$22.5<i<23.5$. 
More details about the photometric redshift calculation can be found in
the explanatory
document\footnotemark\footnotetext{
ftp://ftpix.iap.fr/pub/CFHTLS-zphot-T0007/cfhtls\_wide\_T007\_v1.2\_Oct2012.pdf}
.

With all the above data available, we can produce again the stacked
overdensities, but this time in three-dimensional space. For calculating 
the expected field objects for both X-ray sources and optical galaxies, we 
again use the same criteria and the same catalogs in a $\sim$2 deg$^2$ field.
This is 
very important, especially for the X-ray counterpart's photometric 
catalog that is not complete, so 
that the overdensity measurements would not be affected.  

Finally, using the SDSS\footnote{http://www.sdss3.org/} and CFHT databases, we
visually inspected  all
the counterparts of the X-ray point-like 
sources within 4$r_{500}$ (or 5$r_{500}$ for the
high-$z$ sample) of every cluster. Our aim was to combine the
available redshifts and images, to provide a more reliable list of
cluster members and background/foreground objects, by investigating 
the morphology of cluster members, and assessing the probability of 
sources with no redshift of also being cluster members. Visual inspection
is the only tool we have for the fraction of our sources that do not
have photometric redshifts, either because there was no counterpart
found or because the available photometric bands were less than
four. However, by comparing them to the rest of the population we get
hints about their redshift and their candidacy as
cluster members. In fact, most of these appear to be blue
pointlike sources that are very similar to the spectroscopically confirmed
background sources. The
rest are either very faint or with no counterpart, and it seems that
they are even less likely to be correlated with the clusters (see also relative 
discussion in Ehlert et al. 2014). All our results can be found in Table 1 and
2.

\section{Results}

\subsection{X-ray point source overdensity}

Using the methodology of \S2.2, we calculate the X-ray point source
overdensities of all our sample clusters in annuli up to 5$r_{500}$
(6$r_{500}$ for the high-$z$ sample) where we expect to have reached the
field density. However, the number of sources found in each cluster is
small, especially in the first annulus, which not only is the smallest
one, but is also the one that includes the extracted core (25\% of its area). To
address this issue and derive more robust results, we stacked all
clusters of each subsample (high-$z$ and low-$z$ separately), calculating
the total overdensity for each annulus as described in \S2.5. 
Nevertheless, in what follows we also present the results of all
clusters individually, while we list all data in Tables 1 and 2. 

In Fig. 4 we present the stacking results of the
low-$z$ clusters (left panel) and of the high-$z$ clusters (right
panel). We can see that for the low-$z$ (high-$z$) clusters, 
in the first (and second) annulus, the X-ray overdensity is high
and it drops steeply in the second (third) annulus. However, this behavior is
reversed in the next annuli, where the overdensity rises
again until it drops and converges to the background zero level in
the fifth (sixth) annulus.
\begin{figure*}
\resizebox{9cm}{9cm}{\includegraphics{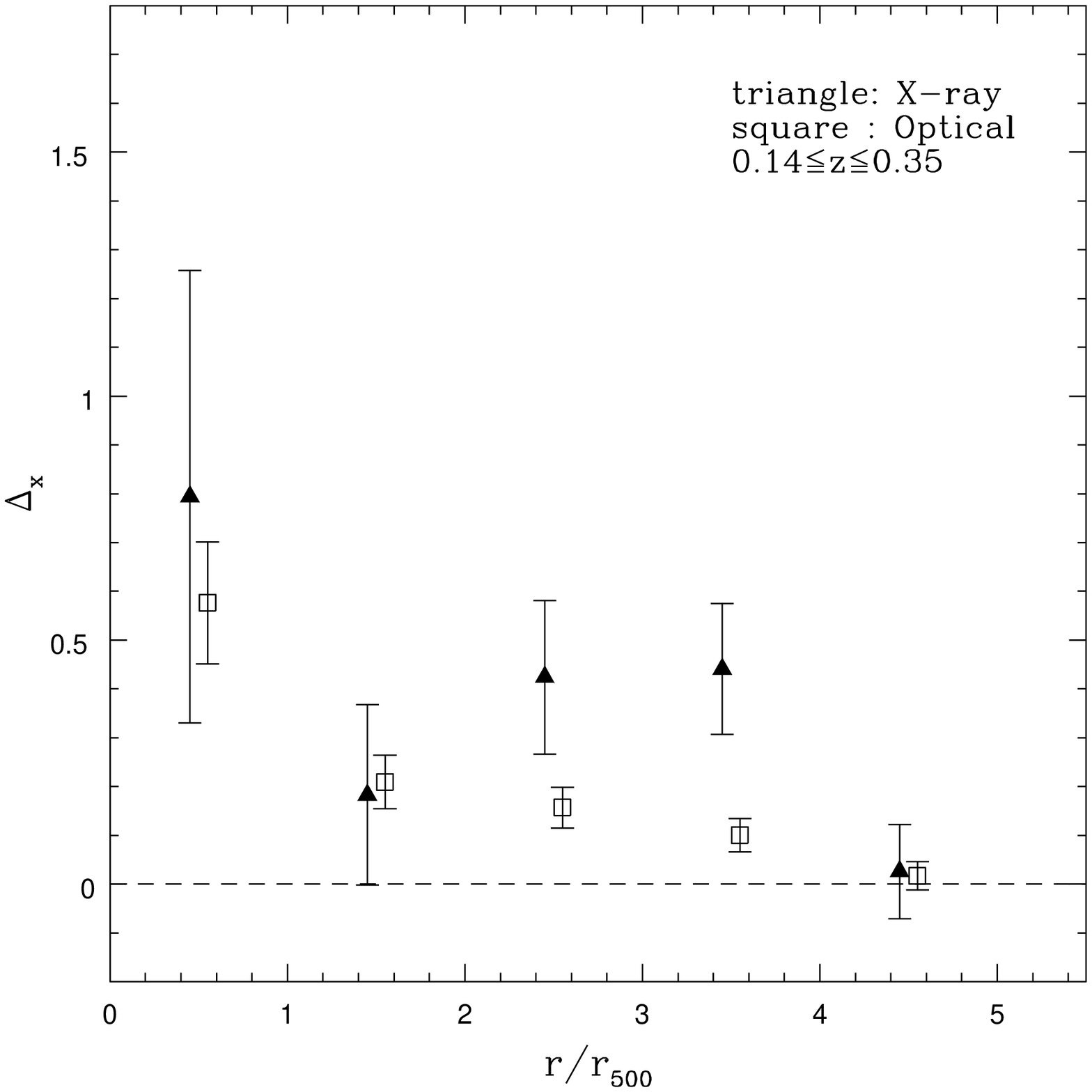}} \hfill
\resizebox{9cm}{9cm}{\includegraphics{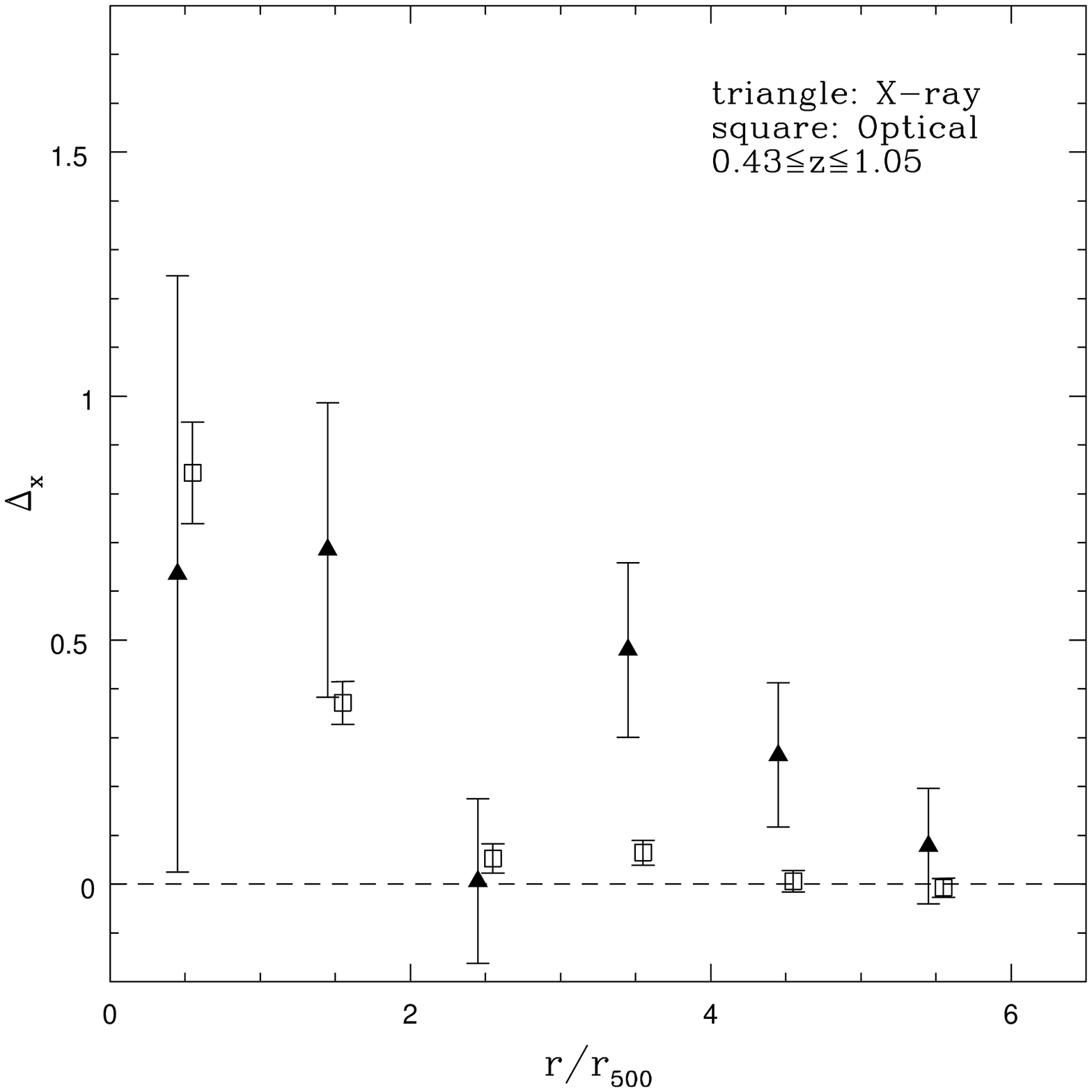}}
\caption{Stacked optical galaxy overdensity (open squares) vs 
X-ray point source overdensity (triangles) as a function of projected
cluster-centric distance. Uncertainties are
1$\sigma$ Poissonian errors of the weighted number of objects.
{\em Left Panel:} Low-$z$ cluster sample. 
{\em Right Panel:} High-$z$ cluster sample.}
\end{figure*}
The X-ray point source excess in the first annulus of both the low and the
high-redshift samples 
is to be expected, since as already mentioned, it is reported in numerous
previous 
studies (e.g., Cappi et al. 2001, Molnar et al. 2002, Johnson et al. 2003,
D'Elia et al. 2004, Gilmour et
al. 2009, Melnyk et al. 2013). The extension of the excess in the second
annulus 
for the high-$z$ sample could be due to
intrinsic differences of the two sets of clusters
or evolutionary effects in the dynamics of the clusters.

In addition, the overdensity \textquotedblleft bump" at larger radii is
statistically significant and is present in
both our cluster samples that are completely independent. The fact
that the rise is not appearing at the same scale for both samples of
clusters could be again due to the above-mentioned differences in the two.

This excess has been reported in previous studies (Ruderman \& Ebeling 2005;
Fassbender et al. 2012) 
and has been attributed to an infalling population of galaxies in the outskirts
of the clusters
that interacts and merges, producing an overdensity of X-ray AGN in the
area. In addition, Haines et al. (2012) compared infalling 
with virialized populations and concluded the same.
Although the analysis of Ehlert et al. (2014) 
stops at 2.5$r_{500}$, the start off the \textquotedblleft bump" is already
apparent
after 2$r_{500}$, but the authors do not comment on that 
assuming that the cluster X-ray source density 
converges to the expected field value at distances of 
$\sim$2$r_{500}$. In contrast, Gilmour et al. (2009) argue that 
any X-ray point source overdensity found at large radii is due to
additional clusters in or near the field of view, which may also
contain AGN and probably also contributes to the enhancement of background
AGN. This surplus is confirmed for our low-redshift
clusters at a smaller radial distance compared to previous works, i.e. 
for the current study approximately between 1 and 2 Mpc but between 2 and 3 Mpc
for 
rich clusters. Nevertheless, our high-$z$ clusters 
are more comparable to rich clusters since their excess is found after 1.5 Mpc.
This comparison to other studies should be considered with caution 
owing to the different methods used for the stacking, i.e. stacking the same
radii in arcmin, Mpc or $r_{500}$. 
Nevertheless, the results appear to be expected 
considering that, on physical scales, the outskirts of poor and moderate 
clusters should be closer to the center of each cluster. This difference is 
already present even between our two samples that consist of clusters with
different
\textquotedblleft richness" (see Tables 1 \&2).

\subsection{X-ray point source versus optical galaxy overdensity}
\subsubsection{Stacking analysis}

As a next step it is essential to compare the already found
X-ray point source overdensities to the 
optical galaxy overdensities. Following the methodology presented in
\S2.4, we calculate the stacked overdensity of optical galaxies 
for the clusters that fall within the CFHTLS area (twelve of the low-$z$ and
eight
of the high-$z$ sample; $\sim60\%$ of each). We do not expect 
that the optical galaxy overdensity profile would vary if we had added
the missing clusters to its calculation. Nevertheless, to verify that 
the comparison between X-ray and optical overdensity is not affected, we 
also limited our X-ray analysis to only the clusters covered by the CFHTLS
and reached the same conclusions. 
\begin{figure*}
\resizebox{9cm}{9cm}{\includegraphics{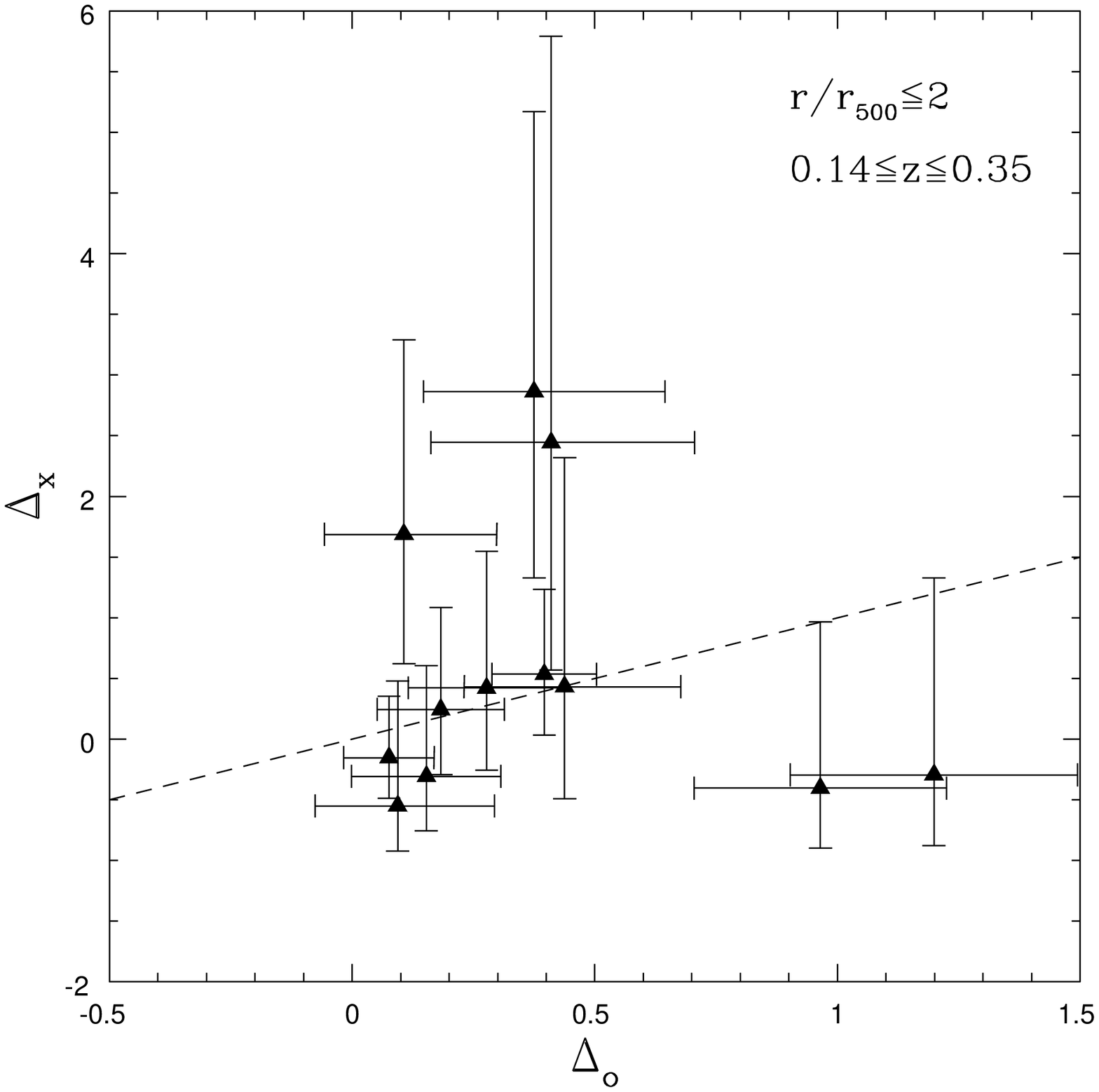}}\resizebox{9cm}{9cm}{
\includegraphics{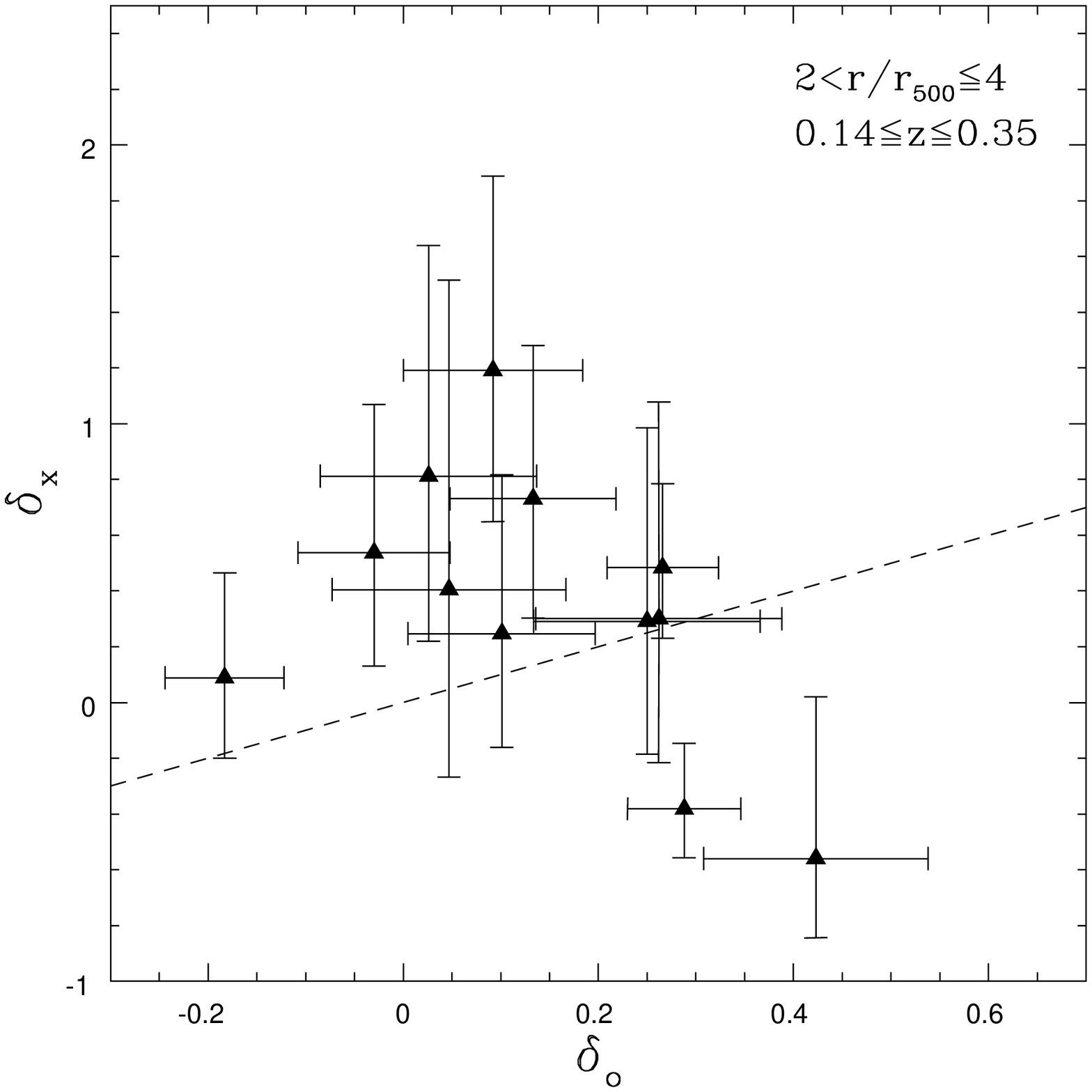}}
\caption{Optical galaxy overdensity vs X-ray point source overdensity
  for each low-$z$ cluster individually. The dashed line corresponds to 
$\Delta_x=\Delta_o$. Uncertainties are Poisson 1$\sigma$ errors for
small numbers (Gehrels 1986).}
\end{figure*}
\begin{figure*}
\resizebox{9cm}{9cm}{\includegraphics{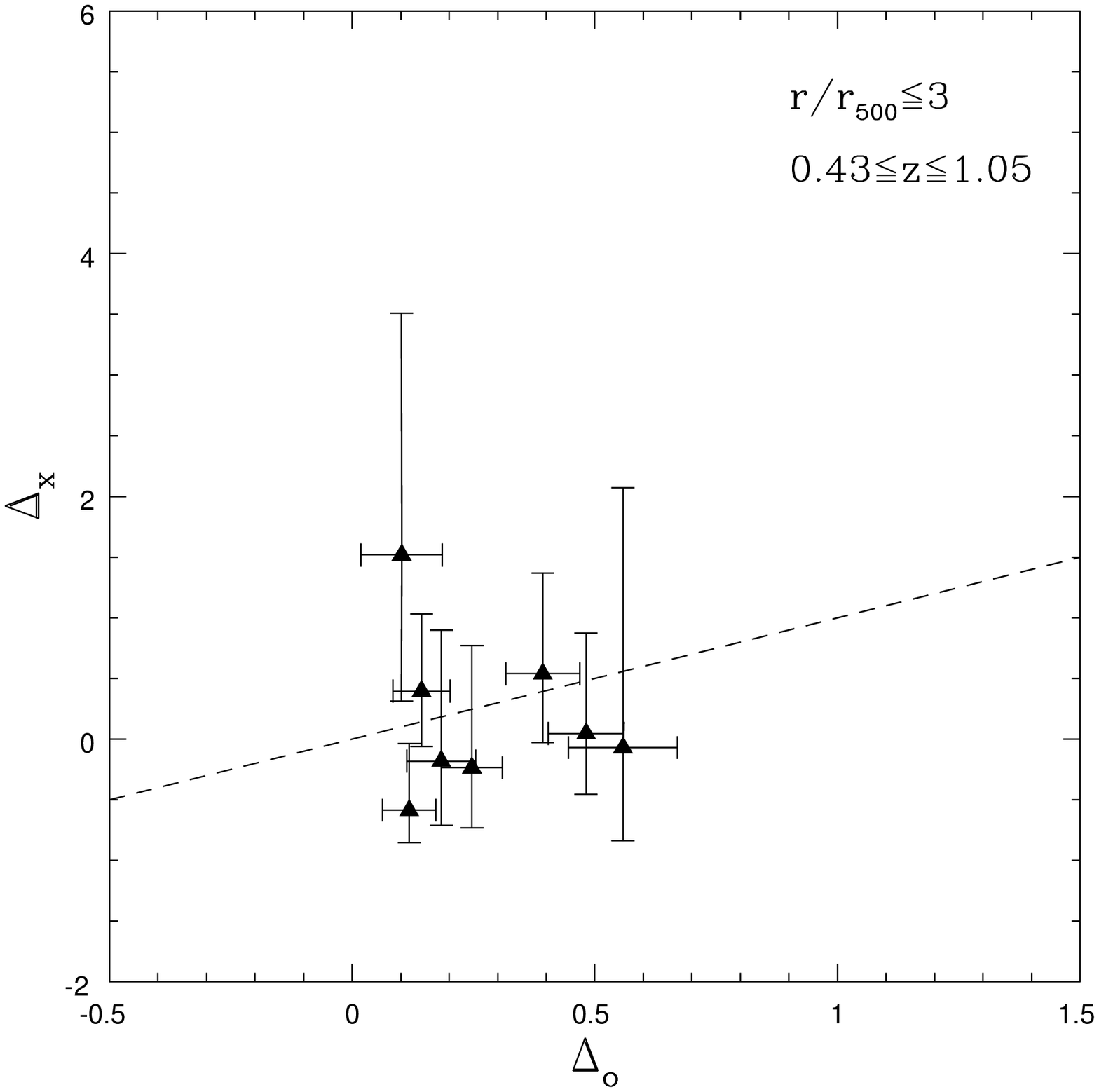}}\resizebox{9cm}{9cm}{
\includegraphics{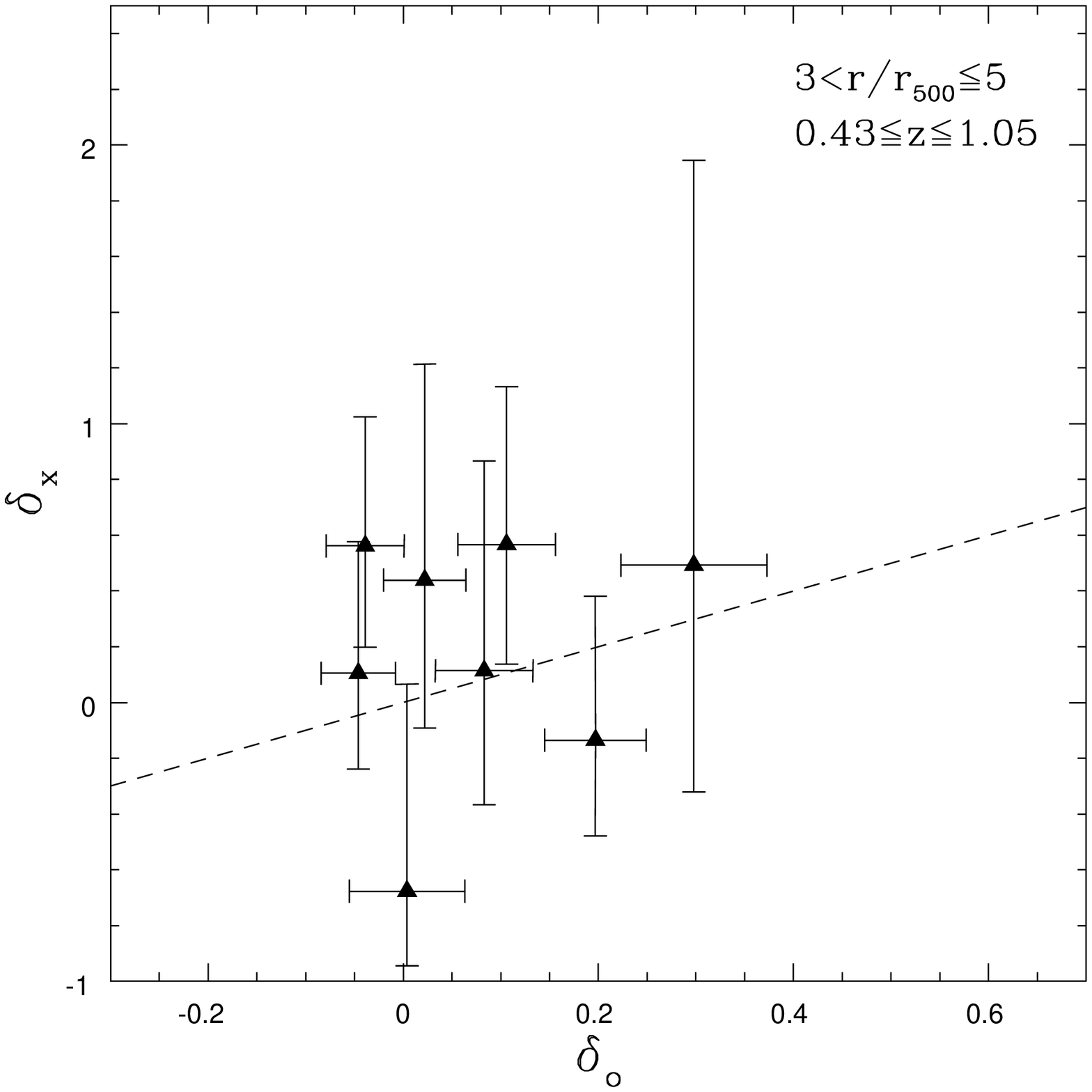}}
\caption{Optical galaxy overdensity vs X-ray point source overdensity
  for each high-$z$ cluster individually. The dashed line corresponds to 
$\Delta_x=\Delta_o$. Uncertainties are Poisson 1$\sigma$ errors for small
numbers (Gehrels 1986).}
\end{figure*}

Therefore, in Fig. 4 we overplot the
optical profile to the X-ray overdensity data. 
For the low-$z$ sample (left panel of
Fig. 4) the X-ray overdensity profile
is consistent with the corresponding optical one in the first two
annuli. Especially in the first one even a small X-ray excess can be seen, but 
it is within the 1$\sigma$ errors and thus not statistically significant. 
Similar results are also found for the high-$z$ sample (right panel of
Fig. 4), where the X-ray overdensity exhibits a more prominent excess in the
second bin and 
agrees with the optical galaxy overdensity in the third. Overall,
we conclude that within the first bins the X-ray overdensity is as expected by
the 
respective optical galaxy results. 
This contradicts to the X-ray AGN suppression that many previous studies of rich
clusters have reported (e.g., Koulouridis \& Plionis 2010; Ehlert et
al. 2013; Haines et al. 2012). Nevertheless, we should note once more that 
the current samples consist of intermediate and low-luminosity clusters in which
the triggering and feeding of the AGN may be more favorable than in richer 
clusters. In addition, the BCGs, which in many cases host an AGN, 
are excluded (since the central region is excluded). The extra X-ray AGN 
would increase the X-ray overdensity dramatically,
but we argue that the BCG's path of evolution is very different from the other
cluster galaxies and should be excluded for the purposes of the current
analysis.

Following the steep drop a rise in the X-ray overdensity
occurs again in the next two bins, but does not appear in the optical data.
Then, they both converge to the background level in the respective last annulus
of our analysis.
Thus, this distant X-ray overdensity \textquotedblleft bump", which was
discovered in the previous section of the current study, 
presents statistically significant differences when also compared to the optical
galaxy overdensity
for both our samples. Consequently, we argue that the abundance of X-ray sources
in 
large clustercentric radii around rich clusters is replicated in poor and
intermediate clusters,
although in somewhat smaller physical distances.

\subsubsection{Individual cluster analysis}
After completing the stacked analysis, we would also like to
investigate the behavior of individual clusters, to clarify if the
discrepancies between optical galaxy and X-ray point source
overdensities emanate from the behavior of all the clusters 
in the samples or just a subsample. Only clusters located in the 
CFHTLS region are used in the current analysis, as explained in the 
methodology.

In Fig. 5 we plot the X-ray vs the optical overdensity for the low-$z$
sample up to the second annulus (left panel) and from the third up 
to the fourth (right panel) for each cluster separately. We selected only
those bins in which a positive X-ray overdensity is found, excluding
the last annulus where it drops to zero. 
Judging by the $r_{500}$ values (Table 1) we can estimate that the
2$r_{500}$ radius corresponds to $\sim$1 Mpc radial distance. 
Therefore, for poor clusters, Fig. 5 is analogous to Fig. 1 of
Koulouridis \& Plionis (2010) for rich Abell clusters. We see that for the 
first two bins more than half of our sample clusters are located close
to the dashed line that denotes equality between optical and X-ray 
overdensities, while the rest are on the one or the other side, canceling
out any discrepancy. However, in the next two bins all but
two clusters move above the line (even marginally) and that is the
reason for the discovered X-ray overdensity \textquotedblleft bump".

We also plot the individual cluster overdensities for the
high-$z$ sample in Fig. 6. To trace the same trends and compare 
with Fig. 5, we add the first three annuli in the left-hand panel and the
next two on the right. Qualitatively, the results are similar to what
we found in Fig. 5. The X-ray \textquotedblleft bump" 
discovered again at larger distances from the cluster center seems to
be produced by a rise in the X-ray point source density in almost all
the clusters.
\begin{figure}
\resizebox{8.5cm}{8cm}{\includegraphics{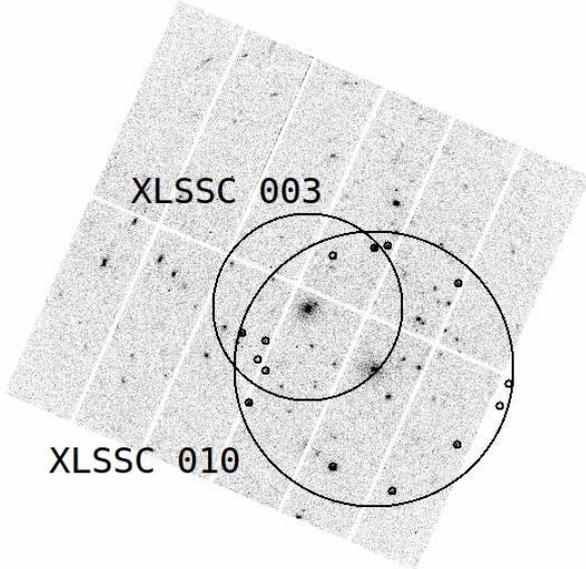}}
\caption{XMM-Newton image of galaxy clusters XLSSC 010 ($z$=0.33) and XLSSC 003
($z$=0.84). The large circles mark the 4$r_{500}$ radius 
of XLSSC 010 and the 5$r_{500}$ radius of XLSSC 003, while the small ones point
toward the detected X-ray point-like sources within the fourth annulus of the 
low-redshift cluster. More than half of the sources seem to be concentrated in
the conjunction of the outer annuli of the two clusters.}
\end{figure}

These results do not agree with the analysis of Gilmour et al. (2009),
where the excess overdensity was attributed solely to additional X-ray AGN
and/or lensing due to 
foreground or background clusters in or near a small fraction of their sample
clusters, 
although the extra lensing seems to be true for some of our low-redshift
clusters that
happen to have a background cluster projected within their area. In Fig. 7 we
present an example of such a case 
where XLSSC 010 is the foreground cluster at redshift $z$=0.33 and XLSSC 003 the
background one at redshift 
$z$=0.84. In the fourth bin of XLSCC 010 about seven sources are expected but 14
found, out of which seven or eight 
are found in the conjunction of the outer bins of both clusters. 

We stress that the results based on the individual and the stacked cluster
analysis are not directly comparable, because 
only clusters that are located in the CFHTLS region are included in the
individual analysis, while we use
the full sample for the X-ray stacked analysis. In addition, no weighting is
performed in the individual analysis, 
thus adding the individual cluster overdensities will not result in the stacked
one.
On the other hand, most of our clusters exhibit high X-ray source overdensities
without any other cluster visible on 
their background. We will probably
need to conduct the same study to the full XXL survey in order to clarify this
issue.  

\subsection{Spatial overdensity analysis}

Having discovered that the X-ray sources exhibit excessive values of
overdensity, not only in the first annuli where they are consistent
with the optical galaxy excess, but also at larger distances where
they are significantly higher, it 
is very important to determine whether this excess is real, i.e., if it can be
attributed
to cluster members.
To this end, we used the available spectroscopic and photometric
redshifts and the methodology of \S2.6. We should note that we are
forced to use larger bin separation in order to have more meaningful results
because the number of clusters 
that have available photometric redshifts is still small. For the
low-redshift clusters we merge the first two bins in one
and the next three bins as well, resulting in a total of two bins, while for the
high-redshift clusters
we use three bins in total.

In Fig. 8 we present the results of the spatial analysis
for the low-$z$ and high-$z$ cluster samples (left and right panels,
respectively). We also plot the results when including 
the ABC region, where the photometric redshifts of the X-ray sources
were calculated with fewer bands. We can see that the uncertainties 
are still very large for the X-ray sources, despite the merging of the annuli.
For the low-$z$ sample,
in the first bin, which corresponds to the first two annuli in the projected
analysis, 
the X-ray and the optical overdensities seem to agree, 
as they also did in the projected overdensity analysis. However, the excess in
the 
second bin, which corresponds to the third and fourth annuli in the projected
analysis, has disappeared. Overall, the number of 
X-ray-selected AGN found within the three cluster bins is exactly the same as
found in the field.
This sharp contrast with our previous projected overdensity results is probably
due to lensing of 
background sources that can affect the projected overdensity analysis but not
the spatial one. 

Our high-$z$ sample exhibits a different behavior. The total overdensity 
of X-ray-selected AGN in the area of clusters is higher than what is expected,
but up to 4$r_{500}$ 
is practically zero (although with large uncertainty) and rises in the last
merged bin.
In addition, the difference with the optical galaxy overdensity is significant
in this last bin, since the 
AGN found are more than double what is expected. Adding the ABC fields only
brings the X-ray 
overdensity closer to the optical in the first merged bin but does not change
the results 
of the other two.

Overall, the trend of X-ray AGN deficiency in
rich galaxy clusters cannot be confirmed for the low-$z$ clusters, 
while for the intermediate ones of the high-$z$ sample, a suppression is
possible in the bins closer to the center, but the results are dubious 
because of small number statistics. Nevertheless, the number of X-ray
counterparts 
that are confirmed in the outer annuli of the high-$z$ sample seems to
corroborate 
previous results that report an excess of X-ray AGN in the outskirts of
clusters. This 
is only true for the richer clusters, probably because 
these are more massive structures, still accreting galaxies that are 
gathering in the outer parts of the clusters and are 
effectively interacting before entering the potential of the cluster and the hot
ICM.
\begin{figure*}
\resizebox{9cm}{9cm}{\includegraphics{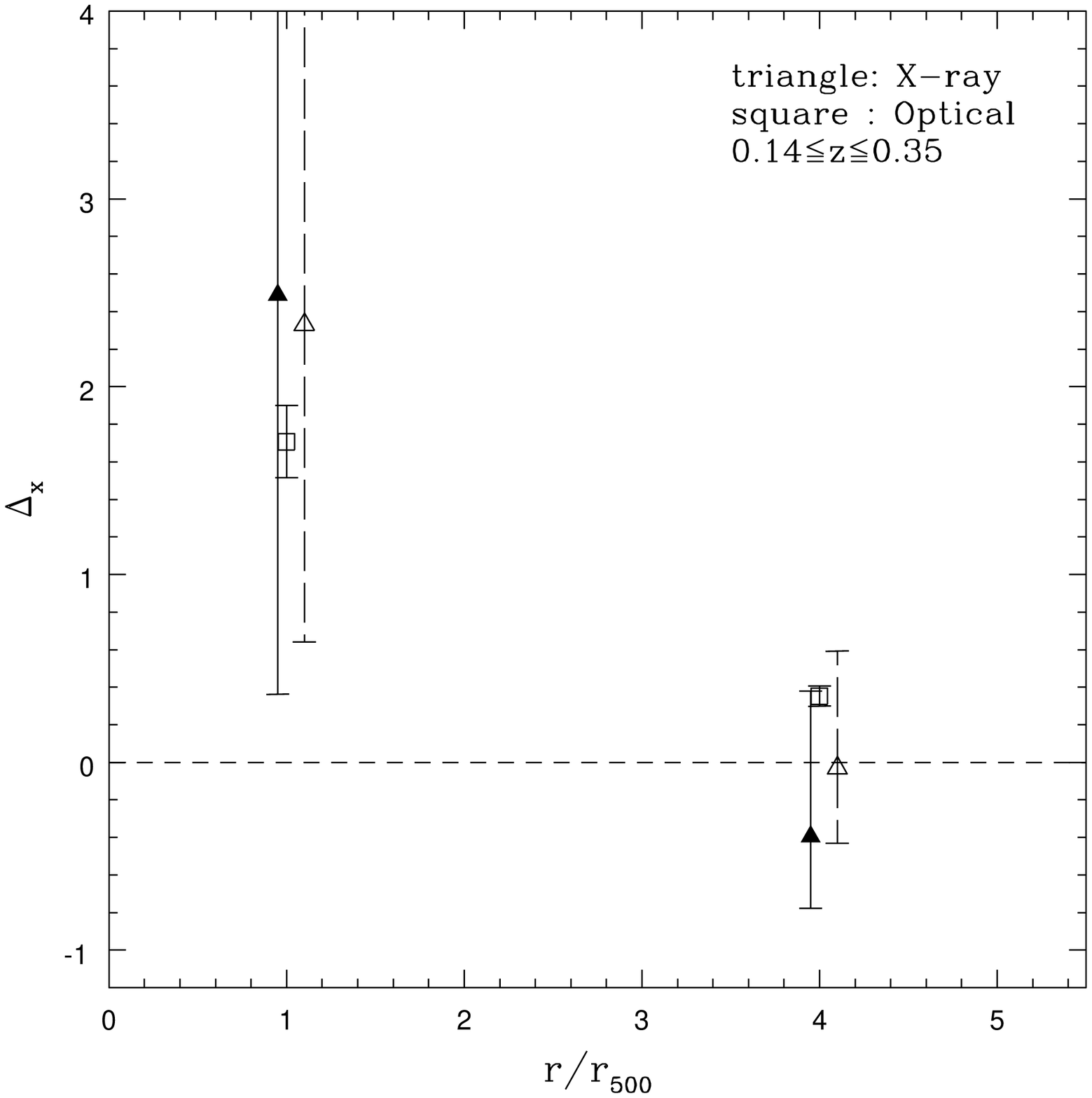}} \hfill
\resizebox{9cm}{9cm}{\includegraphics{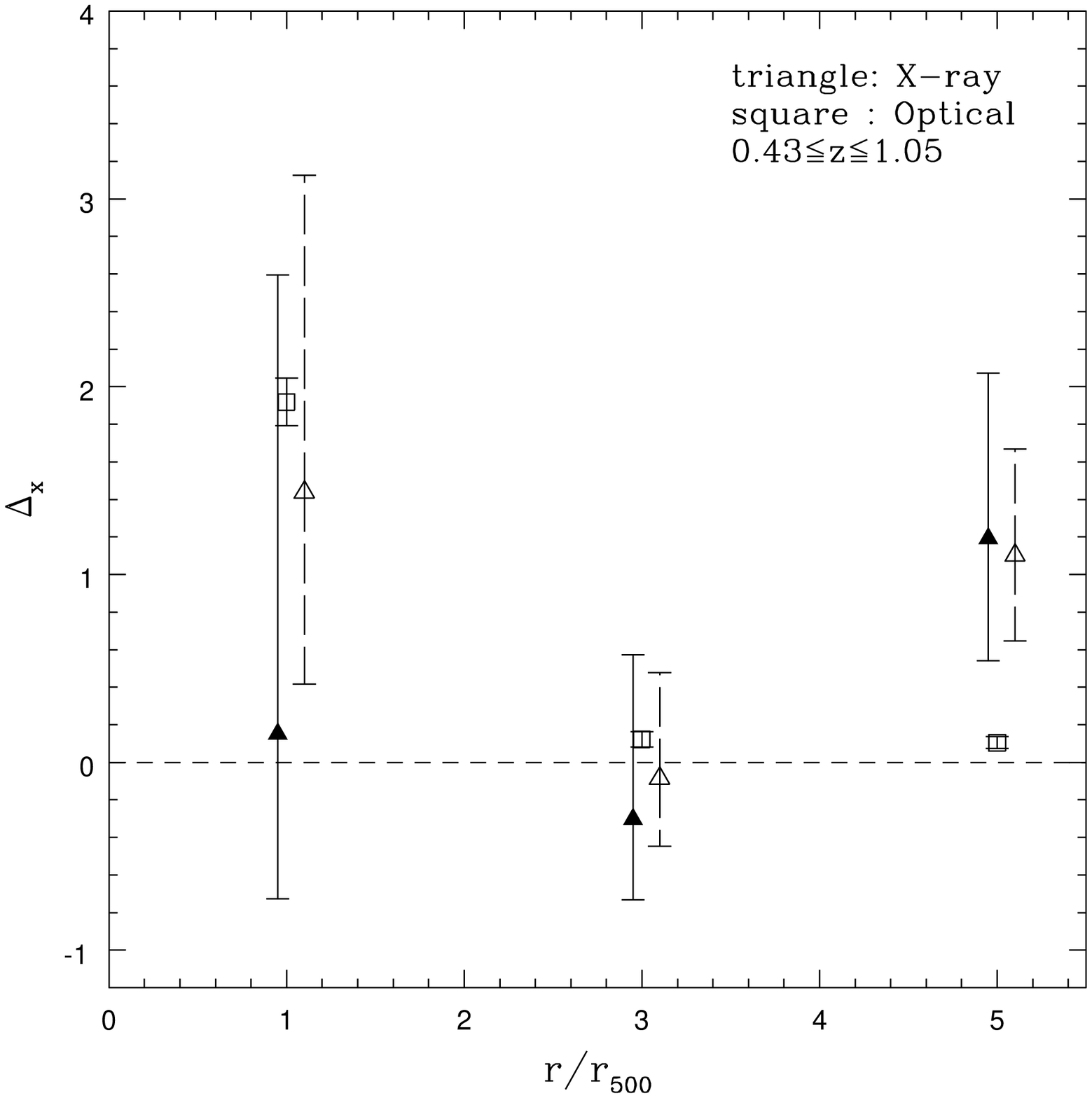}}
\caption{Spatial stacked optical galaxy (open squares) and X-ray
  point source overdensities as a function of projected
  radial distance from the center of the cluster, including only
  clusters in the CFHTLS region (solid triangles) and 
  including clusters in the CFHT ABC supplementary pointings (dashed triangles).
{\em Left panel:} The low-$z$
  cluster sample. {\em Right panel:} The high-$z$ cluster sample. Uncertainties
are Poisson 1$\sigma$ errors for
small numbers (Gehrels 1986).}
\end{figure*}

Considering that redshifts are not available for a large number of X-ray
sources,
we proceed with the visual inspection of all the counterparts of the cluster
member candidates
of the projected overdensity analysis up to the 4th bin (5th bin) for the
low-$z$ (high-$z$) sample.
Our aim is to investigate the morphology of the optical counterparts of the 
confirmed X-ray AGN cluster members and compare them to the ones that lack any
redshift information. We should 
mention, however, that no counterpart is detected in many cases or the
determination of the correct counterpart 
is dubious. 

In Table 1 (columns $(12)-(20)$) we present our results for the low-$z$
sample. The first four columns ($(12)-(15)$), which correspond to the four
$r_{500}$ annuli respectively, contain 
the number of sources that, based on the
redshift of their optical counterparts, are background or foreground (projected)
objects. The following four columns 
($(16)-(19)$ that correspond to the same four $r_{500}$ annuli) instead contain
the sources that
are true cluster members (not included in the projected sources). 
In column $(20)$ we report the sources that lack any
redshift information. Therefore, the sum of the nine columns ($(12)-(20)$) is
the total number of
X-ray sources $N_x$ in all the cluster's annuli, reported in column $(11)$. 
The numbers in parenthesis are the sources with available
spectroscopic redshift and are a subsample of the preceding number of sources,
e.g., 8(3) in column (14) of Table 1
shows that from the total eight sources with redshift information in the third
bin, three have spectroscopic redshifts
and five photometric. Clusters located
in the ABC supplementary fields, where the majority of the
photometric redshifts of the counterparts are of poor quality, are marked with
an x sign in column $(21)$, and 
the respective photometric redshifts are placed within brackets. Table 2
includes the high-z sample, with the only 
difference that the number of columns that correspond to the cluster's annuli
are in this case five (see \S 3.1).

From the total of 274 sources in the low-$z$ clusters, $\sim$30\% have
spectroscopic redshifts, which
rises to 35\% if we exclude the ABC fields. On the
other hand, objects with no redshift information are the 22\% of the
total, but that percentage drops to 10\% when excluding the ABC
fields. The majority of the non-redshift X-ray sources are blue
point-like objects. These counterparts are abundant in our sample, and
when redshift is available, they can securely be classified as 
background QSOs. We have no reason to believe that
these sources belong to any of our clusters. 
Apart from those, a large fraction of non-redshift objects do not have
any counterpart. Finally, 
only a small number of objects look like faint normal galaxies that
may or may not be cluster members, while the reported redshift of a few others
seems improbable. In Tables 1 \& 2 we included or excluded such objects
accordingly.
Judging from the results of objects
with available redshift, we argue that the probability of 
sources with no available redshift to be cluster members is very low, especially
for the
clusters that fall in the CFHTLS region. Thus, we argue that not considering
them 
in the spatial analysis of our samples does not alter our results. 

We conducted the same analysis for our high-redshift clusters, and
found that from the total of 233 sources 17\% have spectroscopy, which
rises to 30\% if we exclude the ABC fields.
On the other hand, objects with no
redshift information are the $\sim$20\% of the total, similar to the low-z
sample.

SDSS images of the optical counterparts 
of the 15 X-ray AGN located 
within the first $r_{500}$ annulus of the 19 low-redshift ($0.14\leq z\leq0.35$)
clusters can be found in Fig. 9. Only XLSSC 025(2) is
a confirmed cluster member, while most of the rest counterparts are blue
point-like background objects. 
Also, some examples of CFHT i-band images of X-ray source counterparts, with
photometric or spectroscopic redshifts that indicate 
they are true cluster members, can be seen in Fig. 10. Their extended morphology
is a further evidence that they are not projected background QSOs.

\begin{figure*}
\resizebox{18cm}{18cm}{\includegraphics{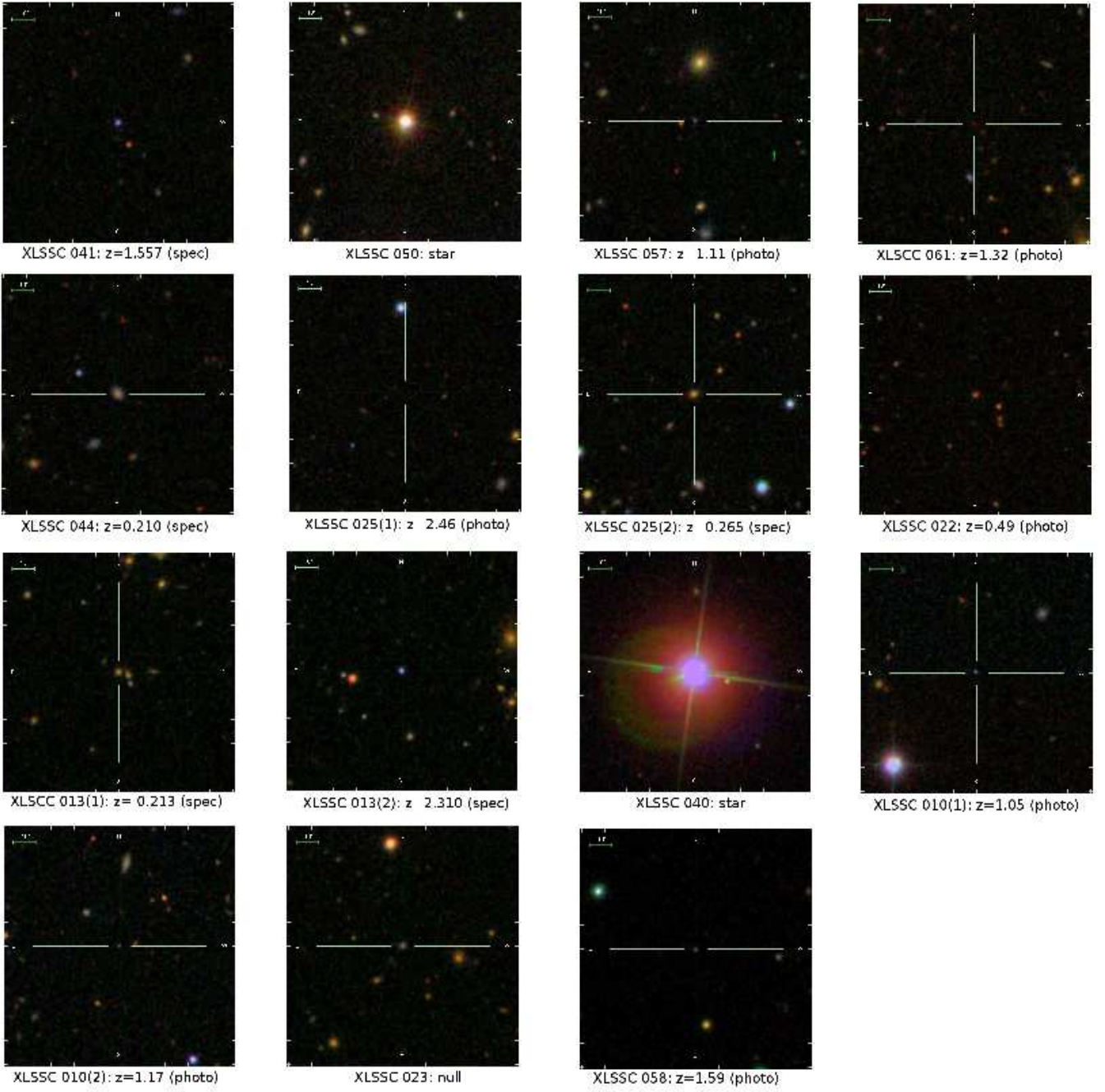}}
\caption{SDSS images of the 15 sources located within $r_{500}$ from the center
of the 19 low-redshift ($0.14\leq z\leq0.35$) clusters. Only XLSSC 025(2) is
a confirmed cluster member. In the parenthesis we indicate if the redshift is
spectroscopic or photometric. The scale can be seen
In the upper left corner of each image.}
\end{figure*}

\begin{figure*}
\resizebox{18cm}{9cm}{\includegraphics{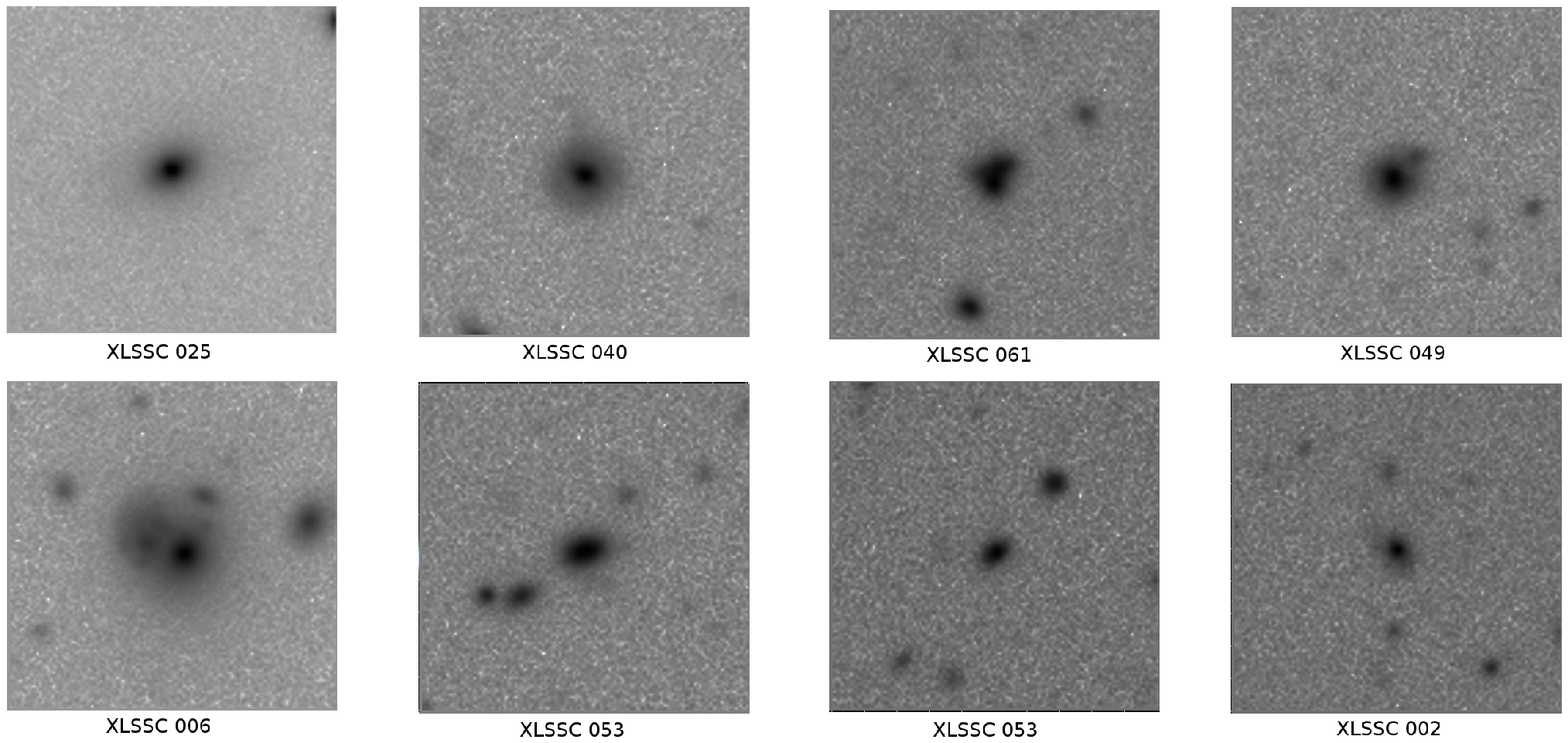}}
\caption{Examples of CFHTLS i-band images of X-ray source counterparts, with
photometric or spectroscopic redshifts indicative of them 
being true cluster members. The dimensions of each image is $30''\times30''$}
\end{figure*}

\section{Conclusions}
We conducted a statistical study of 33 clusters of poor and moderate richness,
within the XMM-LSS
field that covers $\sim$20$\%$ of the XXL survey, by comparing 
the density of X-ray sources within multiples of the $r_{500}$
radius with the expected field density, calculated 
from the $\log N - \log S$ for the same area. We compared this projected
overdensity
with the respective optical
galaxy overdensity in an attempt to estimate the suppression or the enhancement
of X-ray-selected AGN. 
In addition, we calculated the spatial overdensities, using the
available spectroscopic and photometric redshifts in an attempt to
identify and quantify the true cluster members and explain the results 
of our previous projected analysis.   

The conclusions that can be drawn from the above analysis are the following:

$ \bullet$ The projected analysis of X-ray versus optical overdensity
within the two central $r_{500}$ annuli, corresponding to $\sim 1$ Mpc radius, 
results in a strong positive signal
showing that the environment of the low and moderate X-ray-luminosity clusters
of our
samples does not suppress the X-ray AGN activity.
This result is in sharp contrast to the outcome of
many studies of rich clusters (e.g., Koulouridis \& Plionis 2010; Ehlert et
al. 2013; Haines et al. 2012), which implies
that lower richness cluster environments do not suppress X-ray AGN activity.
Interestingly, in even lower density environments (galaxy groups), an
enhancement 
of X-ray AGN may be present (Melnyk et al. 2013).

$ \bullet $ After calculating the projected overdensities at large radial
distances 
from the center of the cluster ($3^{rd}-5^{th}$ bins, corresponding to $\sim
1.5$-3 Mpc,
depending on the redshift), a significant rise in the X-ray source overdensity
is observed.
This excess has also been reported in previous studies (Haines
et al.2012, Fassbender et al. 2012) and has been attributed to an
infalling population of galaxies from the outskirts of the clusters
that interact and merge, producing the observed overdensity of X-ray AGN.
This surplus is confirmed for both our low- and high-redshift clusters. 

$ \bullet $  Using spectroscopic and photometric redshifts, 
we discovered that the X-ray \textquotedblleft bump" at a large radial distance 
vanishes completely from the poor low-$z$ sample, and we argue therefore that
this density excess may be produced by flux boosting of background sources 
due to gravitational lensing, sometimes even enhanced by additional background 
galaxy clusters along the same line of sight. 
On the other hand, a high X-ray source overdensity persists 
in the last annulus of the moderate X-ray luminosity high-$z$ sample, implying 
that for intermediate \textquotedblleft -richness" clusters, additional 
triggering of X-ray AGN in the outskirts is still possible.  
 
In a nutshell, the projected overdensity analysis produces 
statistically significant results, but at the same time these results 
are contaminated by projection effects of background-lensed QSOs.
On the other hand, although the spatial analysis performed is free of these
effects, 
it is not able to reach definite results owing to the small numbers involved,
making it necessary to study larger samples of galaxy clusters. 
Especially the area included in the 
annuli closer to the cluster center is so small that does not allow us
to reach any definitive conclusions about the suppression of X-ray AGN. The
stacking of 
clusters proves very useful, but splitting the total sample into two redshift
subsamples
again reduces the numbers greatly. However, the division is crucial since not
only 
do we select a population of more X-ray luminous clusters in higher redshifts, 
but we can also detect only higher luminosity X-ray AGN.

We should stress that
the large contiguous area of the XMM-LSS has allowed us to study
the overdensity
of X-ray AGN 
within large radial distances from the cluster center for the first time. This
proved to be  
essential for exploring the relation between the dense environment of clusters 
and the  X-ray AGN activity in detail.  
To fully understand this relation, we need to trace its evolution
as a galaxy approaches the cluster's gravitational potential, enters the hot
ICM, and crosses the cluster. 
At the same time, we need to disentangle irrelevant effects such as the
gravitational lensing of background sources,
probably enhanced by the presence of additional clusters along the line of
sight. 
A photometric variability study of these sources may also shed some light on
whether the lensing amplification 
could be due to micro-lensing and/or convergence by matter in the clusters.

We believe that the analysis of the full XXL field, which is almost five times
larger than the XMM-LSS 
(reaching 50 deg$^2$), 
together with a detailed spectroscopic follow-up of the optical 
counterparts of all X-ray point sources, detected in the XXL clusters, 
may provide reliable and robust results as to the origin (true enchancement, 
lensing, presence of background clusters, etc) of the excess X-ray sources
detected in the outer $\sim 3-5 r_{500}$ annuli 
of either low and high-redshift clusters.
\acknowledgements
We would like to thank the
anonymous referee who provided insightful comments and suggestions. 
EK acknowledges fellowship funding provided by the Greek General Secretariat of Research
and Technology in the framework of the programme Support of Postdoctoral
Researchers, PE-1145.
This work is based on observations obtained with XMM-Newton, an ESA science
mission with instruments and contributions directly funded by ESA Member States
and the USA (NASA). Funding for SDSS-III has been provided by the Alfred P. Sloan Foundation, the
Participating Institutions, the National Science Foundation, and the U.S.
Department of Energy Office of Science. The SDSS-III web site is
http://www.sdss3.org/. This work is based on observations obtained with MegaPrime/MegaCam, a joint project of CFHT
and CEA/IRFU, at the Canada-France-Hawaii Telescope (CFHT) which is operated by
the National Research Council (NRC) of Canada, the Institut National des Science
de l'Univers of the Centre National de la Recherche Scientifique (CNRS) of
France, and the University of Hawaii. This work is based in part on data
products produced at Terapix available at the Canadian Astronomy Data Centre as
part of the Canada-France-Hawaii Telescope Legacy Survey, a collaborative
project of NRC and CNRS.

\newpage

\begin{landscape}
\begin{table}
\begin{minipage}{200mm}
\caption{Low-redshift sample} 

\tabcolsep 3 pt
\begin{tabular}{lcccccccccc|cccc|cccc|cc} \\ \hline

name & $z$ & kT & $L_x, bol$  & $r_{500}$& AGN-$L_x$ & $\Delta_x$  & $\Delta_o$
&$\delta_x$ &$\delta_o$  &N$_x$  &
 N$_{x1}$ & N$_{x2}$ & N$_{x3}$ & N$_{x4}$ & N$_{x1}$ & N$_{x2}$ & N$_{x3}$ &
N$_{x4}$ & no-z & ABC \\ 
{\em (1)}&{\em (2)}&{\em (3)}&{\em (4)}&{\em (5)}&{\em (6)}&{\em (7)}&{\em
(8)}&{\em (9)}&{\em (10)}&{\em (11)}&{\em (12)}&{\em (13)}&{\em (14)}
&{\em (15)}&{\em (16)}&{\em (17)}&{\em (18)}&{\em (19)}&{\em (20)}&{\em (21)}\\
\hline
\multicolumn{10}{l}{}&&\multicolumn{3}{l}{\bf Projected
sources}&&\multicolumn{3}{l}{\bf Cluster members}&&&
\\
XLSSC 041 & 0.140 & 1.6 &  2.74e+43&   509  & 1.0e+42 &  2.444  &  0.410 & 0.405
 & 0.047  & 7   &   1(1) & 1(1) & 2(2)& 2(2)&-&1&-&-&-& \\
XLSSC 050 & 0.140 & 3.3 &  9.19e+43&   769  & 1.0e+42 &  0.878  &    -   & 0.896
 &   -    & 17  &   [1](1) & [3] &[8](5)&[4](2)&-&-&-&-&1& x \\
XLSSC 057 & 0.155 & 2.0 &  3.07e+43&   570  & 1.0e+42 &  0.431  &  0.438 &-0.560
 & 0.423  & 4   &   1  &1(1)&-&2&-&-&-&-&-&  \\
XLSSC 077 & 0.200 & 1.6 &  1.20e+43&   469  & 1.0e+42 &  0.396  &    -   &-0.787
 &   -    & 3   &   -  &[1](1)&-&-&-&-&-&-&2& x \\
XLSSC 039 & 0.231 & 1.0 &  6.22e+42&   349  & 1.0e+42 &  0.029  &    -   &-0.057
 &  -     & 4   &   -&-&-&-& -&-&-&-&4& x\\
XLSSC 061 & 0.257 & 1.8 &  2.84e+43&   499  & 1.0e+42 &  1.687  &  0.107 & 1.192
 & 0.092  & 22  &   1  &5(1)&8(2)&8&-&-&-&-&-&  \\
XLSSC 044 & 0.262 & 1.2 &  1.05e+43&   388  & 1.0e+42 & -0.295  &  1.119 & 0.301
 & 0.262  & 7   &   1(1)&-&1(1)&2(1)&-&-&-&-&3&  \\
XLSSC 025 & 0.266 & 2.1 &  4.94e+43&   541  & 1.0e+42 &  0.422  &  0.277 & 0.538
 &-0.030  & 18  &   1&2(1)&8(3)&5(2)&1(1)&-&-&-&1&  \\
XLSSC 051 & 0.279 & 1.4 &  7.02e+42&   418  & 1.0e+42 &  0.131  &    -   & 0.926
 &   -    & 13  &   -&[1](1)&-&[2](1)&-&-&-&-&10&x \\
XLSSC 022 & 0.293 & 2.0 &  6.59e+43&   514  & 1.0e+42 & -0.308  &  0.153 & 0.731
 & 0.133  & 18  &   1&1&4(1)&11(5)&-&-&-&-&1&  \\
XLSSC 028 & 0.296 & 1.2 &  1.00e+43&   369  & 1.0e+42 & -1.000  &    -   &-0.369
 &   -     & 3   &   -&-&-&[2](1)&-&-&[1]&-&-&x  \\
XLSSC 008 & 0.299 & 1.3 &  1.09e+43&   392  & 1.0e+42 & -0.404  &  0.965 & 0.291
 & 0.250  & 8   &   -&1&1(1)&5(2)&-&-&-&-&1&  \\
XLSSC 013 & 0.307 & 1.2 &  1.44e+43&   367  & 1.0e+42 &  2.863  &  0.375 & 0.812
 & 0.026  & 15  &   2(2)&4(3)&1&6(4)&-&-&-&-&2&  \\
XLSSC 040 & 0.320 & 3.9 &  2.02e+43&   775  & 1.0e+42 & -0.153  &  0.076 & 0.484
 & 0.266  & 40  &   1(1)&3(1)&14(6)&17(8)&-&1&-&-&4&  \\
XLSSC 018 & 0.324 & 1.5 &  1.21e+43&   424  & 1.0e+42 & -0.552  &  0.094 & 0.247
 & 0.101  & 10  &   -&1(1)&5(1)&3&-&-&-&-&1&  \\
XLSSC 010 & 0.331 & 2.5 &  5.29e+43&   581  & 1.0e+42 &  0.731  &    -   & 0.817
 &   -    & 30  &   [2]&[1]&[2](2)&[2](1)&-&-&-&-&23&x \\
XLSSC 023 & 0.328 & 1.9 &  3.63e+43&   496  & 1.0e+42 & -0.003  &    -   & 0.357
 &   -    & 16  &   -&[2]&[4](1)&[6](2)&-&-&[1]&-&3&x  \\
XLSSC 058 & 0.333 & 2.4 &  1.63e+43&   564  & 1.0e+42 &  0.244  &  0.183 & 0.089
 & -0.183  & 19  &   1&3(2)&6(3)&7(2)&-&-&-&-&2&  \\          
XLSSC 056 & 0.350 & 3.5 &  1.24e+44&   722  & 1.1e+42 &  0.535  & 0.396  &
-0.381 & 0.288   &  20   &   -&6(1)&6&4(1)&-&-&-&1&3&  \\                
\hline

\hline
\end{tabular}
\tablefoot{{\em (1)} original name in the XMM-LSS database, {\em (2)}, redshift,
{\em (3)} cluster's temperature in keV, 
{\em (4)} cluster's luminosity in erg s$^{-1}$, {\em (5)} $r_{500}$ in Mpc, {\em
(6)} point source's lower X-ray luminosity limit in erg s$^{-1}$ 
in the [0.5-2] keV band, 
{\em (7)} X-ray point source overdensity up to 2$r_{500}$, {\em (8)} optical
galaxy overdensity up to 2$r_{500}$, 
{\em (9)} X-ray point source overdensity from 2 to 4$r_{500}$, {\em (10)}
optical galaxy overdensityfrom 2 to 4$r_{500}$,
{\em (11)} total number of X-ray sources up to 4$r_{500}$, 
{\em (12)-(15)} number of projected X-ray sources in annuli 1 to 4,
respectively; 
the parentheses denote the spectroscopically confirmed sources,
{\em (16)-(19)} as for {\em (12)-(15)} but for true cluster members (projected
sources and cluster members are disjoint), {\em (20)} sources with no redshift,
{\em (21)} when marked the clusters 
are located in the CFHT ABC fields (see Fig. 1), where redshifts were calculated
using only 4-6 photometry bands.\newline 
Numbers in brackets denote sources in the ABC supplementary fields.} 
\end{minipage}
\end{table}
\end{landscape}

\begin{landscape}
\begin{table}
\begin{minipage}{220mm}
\caption{High-redshift sample} 

\tabcolsep 3 pt
\begin{tabular}{lcccccccccc|ccccc|ccccc|cc} \\ \hline

name & $z$ & kT & $L_x, bol$  & $r_{500}$& AGN-$L_x$ & $\Delta_x$  & $\Delta_o$
&$\delta_x$ &$\delta_o$  &N$_x$  &
 N$_{x1}$ & N$_{x2}$ & N$_{x3}$ & N$_{x4}$ & N$_{x5}$ & N$_{x1}$ & N$_{x2}$ &
N$_{x3}$ & N$_{x4}$ & N$_{x5}$& no-z & ABC \\ 
{\em (1)}&{\em (2)}&{\em (3)}&{\em (4)}&{\em (5)}&{\em (6)}&{\em (7)}&{\em
(8)}&{\em (9)}&{\em (10)}&{\em (11)}&{\em (12)}&{\em (13)}&{\em (14)}
&{\em (15)}&{\em (16)}&{\em (17)}&{\em (18)}&{\em (19)}&{\em (20)}&{\em
(21)}&{\em (22)}&{\em (23)}\\
\hline
\multicolumn{10}{l}{}&&\multicolumn{4}{l}{\bf Projected
sources}&&\multicolumn{4}{l}{\bf Cluster members}&&&
\\
XLSSC 006 & 0.429 & 5.6 & 6.70e44& 0.924 & 1.7e42&  -0.019 &   -  & 0.175&    - 
 &  55   & [2]&[4]&[10](2)&[18](2)&[13](2)&-&-&-&-&[1]&7& x \\
XLSSC 036 & 0.494 & 3.8 & 3.01e44& 0.694 & 2.4e42&  1.501 &   -  & 0.511&    -  
&  45   & [1]&[8](1)&[7]&[11](2)&[7]&-&-&[1]&-&[1]&9&x \\
XLSSC 053 & 0.494 & 3.5 & 3.26e44& 0.498 & 2.4e42&  0.539 & 0.393& 0.567&  0.106
&  20   & -&2&4(2)&3&6(2)&-&1&-&-&1&3&  \\
XLSSC 049 & 0.496 & 2.9 & 5.02e43& 0.583 & 2.4e42&  0.395 & 0.143& 0.562& -0.039
&  27   & -&3(3)&3(3)&8(3)&7(3)&-&-&1&1&-&4&  \\
XLSSC 001 & 0.614 & 3.3 & 3.57e44& 0.591 & 4.0e42& -0.585 & 0.118 & 0.105&
-0.046 &  12   & -&1&-&1&-&-&-&-&-&2&8&  \\
XLSSC 059 & 0.645 & 2.7 & 6.20e43& 0.564 & 4.5e42&  0.046 & 0.078&-0.135&  0.197
&  10   & 1&-&1(1)&1(1)&3(2)&-&-&-&-&1&3&  \\
XLSSC 080 & 0.646 & 1.5 & 3.41e43& 0.352 & 4.5e42&  1.517 & 0.102&-0.677&  0.004
&  5    & 1(1)&-&3(2)&-&1&-&-&-&-&-&-&  \\
XLSSC 076 & 0.752 & 1.3 & 9.63e43& 0.298 & 6.4e42& -0.070 & 0.558& 0.493&  0.298
&  4    & -&-&1&2&-&-&-&-&-&-&1&  \\
XLSSC 002 & 0.772 & 2.6 & 1.83e44& 0.467 & 6.7e42& -0.236 & 0.247& 0.439&  0.022
&  9    & -&1&-&2&2(2)&-&1&-&-&1&2&  \\
XLSSC 003 & 0.835 & 3.2 & 3.34e44& 0.509 & 8.2e42& -0.632 &   -  & 1.326&    -  
&  13   & -&-&[1]&[2]&[3]&-&-&-&[3]&[1]&3&x \\ 
XLSSC 078 & 0.960 & 3.3 & 1.32e44& 0.479 & 1.1e43& -0.182 & 0.114& 0.114&  0.083
&  7    & -&1(1)&1(1)&2&3(1)&-&-&-&-&-&-&  \\
XLSSC 072 & 1.003 & 3.5 & 5.28e44& 0.485 & 1.3e43&  1.766 &   -  & 0.671&    -  
&  13   & [1]&[2]&-&[2]&[1]&-&[2]&-&[1]&[1]&3&x \\
XLSSC 048 & 1.005 & 3.0 & 1.61e44& 0.439 & 1.3e43& -0.440 & -    & 0.459&   -   
&  6    & -&-&[1]&-&[1]&-&-&-&-&[2]&2&x  \\
XLSSC 005 & 1.053 & 2.7 & 1.38e44& 0.400 & 1.4e43&  0.789 &   -  & 0.300&   -   
&  7    & -&[2](2)&-&[2](1)&[1]&-&[1]&-&-&[1]&-&x  \\

\hline
\end{tabular}
\tablefoot{{\em (1)} original name in the XMM-LSS database, {\em (2)}, redshift,
{\em (3)} cluster's temperature in keV, 
{\em (4)} cluster's luminosity in erg s$^{-1}$, {\em (5)} $r_{500}$ in Mpc, {\em
(6)} point source's lower X-ray luminosity limit in erg s$^{-1}$ 
in the [0.5-2] keV band, 
{\em (7)} X-ray point source overdensity up to 3$r_{500}$, {\em (8)} optical
galaxy overdensity up to 3$r_{500}$, 
{\em (9)} X-ray point source overdensity from 3 to 5$r_{500}$, {\em (10)}
optical galaxy overdensity from 3 to 5$r_{500}$,
{\em (11)} total number of X-ray sources up to 5$r_{500}$, 
{\em (12)-(16)} number of projected X-ray sources, in annuli 1 to 5,
respectively; 
the parentheses denote the spectroscopically confirmed sources,
{\em (16)-(20)} as for {\em (12)-(16)} but for true cluster members (projected
sources and cluster members are disjoint), {\em (22)} sources with no redshift,
{\em (23)} when marked the clusters 
are located in the CFHT ABC fields (see Fig. 1), where redshifts were calculated
using only 4-6 photometry bands.\newline 
Numbers in brackets denote sources in the ABC supplementary fields.\newline The
cluster XLSSC 078, although covered by CFHTLS, is excluded from the optical
stacking analysis
because $m_i^*+1>24$.} 
\end{minipage}
\end{table}
\end{landscape}

\end{document}